\documentclass[preprint,3p,11pt,times]{elsarticle}

\usepackage[T1]{fontenc}
\usepackage[utf8]{inputenc}

\usepackage{hyperref}
\hypersetup{
    pdftitle={Impact of cyclostationarity on fan broadband noise prediction},
  pdfauthor={Attila Wohlbrandt, Carolin Kissner, Sebastien Guerin},
  pdfkeywords={Modeling, stochastic sources, fan broadband noise, CFD, CAA, cyclostationarity},%
  colorlinks=false,
  linkcolor=black,
  citecolor=black,
  pdfborder={0 0 0},
  breaklinks=true, 
  naturalnames=false, 
  hypertexnames=true,
  bookmarksnumbered=true,
  bookmarksdepth=subsection,
  pdfstartview={Fit}
}
\usepackage{etoolbox}
\makeatletter
\patchcmd{\@decl@short}{\bbl@info}{\@gobble}{}{}
\patchcmd{\@decl@short}{\bbl@info}{\@gobble}{}{}
\makeatother
\usepackage{scrhack}

\usepackage{chngcntr}
\counterwithout{equation}{section}

\usepackage{amsmath}
\usepackage{amssymb}

\usepackage{multicol}

\usepackage[per-mode = symbol]{siunitx}  

\usepackage{color}
\usepackage{soul}

\newcommand{\bfm}[1]{\boldsymbol{#1}}

\usepackage{xspace}
\newcommand{\Karman}{K\'arm\'an\xspace}

\usepackage[english]{varioref}
	\labelformat{equation}{(#1)}

\usepackage{mathrsfs}
\usepackage{amsbsy}

\DeclareMathOperator\const{const}
\newcommand{\abl}{\text{d}}

\newcommand{\x}{{\bfm x}}

\renewcommand{\u}{{\bfm u}}
\newcommand{\e}{{\bfm e}}

\newcommand{\pAbl}[2]{\frac{\partial #1}{\partial #2}}

\newcommand{\bpm}{\begin{pmatrix}}
\newcommand{\epm}{\end{pmatrix}}
\newcommand{\bit}{\begin{itemize}}
\newcommand{\eit}{\end{itemize}}
\usepackage{mathrsfs}
\usepackage{subfigure}
\usepackage{subfigmat}
\usepackage{pdflscape}
\usepackage{multirow}

\usepackage{enumitem} 

\usepackage{tikz} 
\usetikzlibrary{fit,shapes}
\usetikzlibrary{trees}

\usepackage{wrapfig}

\usepackage{overpic}
\usepackage{xargs}

\usepackage[disable]{todonotes}

\newcommandx{\unsure}[2][1=]{\todo[linecolor=red,backgroundcolor=red!25,bordercolor=red,#1]{#2}}
\newcommandx{\change}[2][1=]{\todo[linecolor=blue,backgroundcolor=blue!25,bordercolor=blue,#1]{#2}}
\newcommandx{\info}[2][1=]{\todo[linecolor=green,backgroundcolor=green!25,bordercolor=green,#1]{#2}}
\newcommandx{\improvement}[2][1=]{\todo[linecolor=violet,backgroundcolor=violet!25,bordercolor=violet,#1]{#2}}
\newcommandx{\missing}[2][1=]{\todo[linecolor=red,backgroundcolor=red!25,bordercolor=red,#1]{#2}}

\journal{Journal of Sound and Vibration}

\begin{document}


\setcounter{page}{1}
\pagenumbering{roman}
 \listoftodos

\clearpage
\begin{frontmatter}
\setcounter{page}{1}
\pagenumbering{arabic}



\title{Impact of cyclostationarity on \\fan broadband noise prediction}

\author{A. Wohlbrandt\footnote{Corresponding Author: attila.wohlbrandt@dlr.de,\\ Telephone +49 30 310006-21, Fax +49 30 310006-39 }}
\author{C. Kissner}
\author{S. Gu\'erin}
\address{Institute of Propulsion Technology, Engine Acoustics Department\\ German Aerospace Center (DLR), M\"uller-Breslau-Str.8, 10623 Berlin, Germany}

\begin{abstract}
One of the dominant noise sources of modern Ultra High Bypass Ratio (UHBR) engines is the interaction of the rotor wakes with the leading edges of the stator vanes in the fan stage.  While the tonal components of this noise generation mechanism are fairly well understood by now, the broadband components are not.  This calls to further the understanding of the broadband noise generation in the fan stage.
This article introduces the cyclostationary stochastic hybrid (CSH) method, which accommodates in-depth studies of the impact of cyclostationary wake characteristics on the broadband noise in the fan stage.  The Random Particle Mesh (RPM) method is used to synthesize a turbulence field in the stator domain using a URANS simulation characterized by time-periodic turbulence and mean flow. The rotor-stator interaction noise is predicted by a two-dimensional CAA computation of the stator cascade. The impact of cyclostationarity is decomposed into various effects investigated separately. This leads to the finding that the periodic turbulent kinetic energy (TKE) and periodic flow have only a negligible effect on the radiated sound power. The impact of the periodic integral length scale (TLS) is, however, substantial. The limits of a stationary representation of the TLS are demonstrated making the CSH method indispensable when background and wake TKE are of comparable level. Good agreement of the CSH method with measurements obtained from the 2015 AIAA Fan Broadband Noise Prediction Workshop are also shown. 


\end{abstract}

\end{frontmatter}


\section{Introduction}



Current and future engines used in civil aviation have large bypass ratios meaning that the fan plays an ever increasing role as a noise source. Fan noise, in particular rotor-stator-interaction (RSI) noise, is one of the most dominant noise sources of an ultra-high bypass ratio (UHBR) engine.  It has the largest contribution during the approach phase and is only surpassed by jet noise during the take-off phase. Its prediction is of an increasing importance for the development of new technologies in light of the overall growth in air traffic and progressively more stringent noise regulations.

The tonal components of this noise generation mechanism have been researched extensively.  As a result, different methods were successfully applied to reduce the tonal RSI noise.  These approaches include reducing the tip circumferential speed to subsonic speeds, using acoustic liners in the engine duct, increasing the rotor-stator gap, choosing certain blade counts to strategically use acoustic cut-off effects, and modifying the blade geometry to e.g. reinforce destructive radial interferences.  Due to a reduction of tonal noise, the relative contribution of the broadband noise has significantly increased.  Hence, a greater understanding of the broadband noise generation mechanism in the fan stage is required to further reduce the RSI noise.

However, the prediction of fan broadband noise is still considered to be a challenge:  On one hand, analytical models are restrictive as they require strong assumptions.  On the other hand, CFD computations that fully resolve turbulent scales are exceedingly demanding in computational resources.   To advance the current understanding of broadband noise generation in a fan stage, methods are needed that are both fast and affordable without being overly restrictive.

Hybrid approaches fill the gap, if it is deemed possible to divide the physical problem into individual phenomena which can be calculated sequentially. This results in a process chain where every task is completed by the most efficient method respectively.  Hybrid approaches can combine numerical, analytical and empirical methods. In broadband noise predictions the two most prominent ones are Large Eddy Simulations (LES) coupled to an acoustic analogy and stochastic methods coupled to a Computational AeroAcoustics (CAA) method. This paper focuses on the latter method. \citet{allan_comparison_2014} compared the two mentioned methods and concluded that the hybrid approach relying on a stochastic method yields satisfactory noise results at a fraction of the cost of LES.

For RSI noise, a hybrid approach can be divided into three main tasks: Firstly, the sound generation mechanisms are modeled either directly or by synthesizing a turbulent field which impinges on the blade row.  Secondly, the sound is propagated considering complex duct geometries and flow. Lastly, the sound is radiated into the far field, i.e. to an observer.  The two last mentioned parts can be realized by a CAA simulation applying the Linearized Euler Equations (LEE). 
The first part, however, has proven to be the crux of the matter. For many years, the only way of modeling the sound sources was to use discrete harmonic gusts to generate the turbulent field~\cite{amiet_high_1976,scott_finite-difference_1995,peake_influence_2004,glegg_panel_2010}.  This method is still in use to model RSI noise. In fact, \citet{lau_effect_2013} have recently investigated the impingement of harmonic gusts in a three-dimensional (3D) CAA simulation to investigate the influence of wavy leading edges.

Aside from this analytically motivated method, two classes of stochastic methods are used to model broadband noise:  the Stochastic Noise Generation and Radiation (SNGR) method and the Random Particle Mesh (RPM) method.

The SNGR methods apply a random set of superposed Fourier modes to realize a target model spectrum, e.g. a von \Karman or a Liepmann spectrum. The SNGR methods can be traced back to the work of \citet{kraichnan_diffusion_1970}, who proposed the theoretical framework, and to \citet{bechara_stochastic_1994}, who was the first to apply it to predict noise generated by free turbulence. \citet{clair_experimental_2013} predicted the effects of wavy leading edges (LE) of isolated airfoils, while \citet{gill_reduced_2014} investigated real symmetric airfoils at zero angle of attack. To predict RSI noise \citet{polacsek_numerical_2015} have presented an approach simulating only one stator vane with periodic boundary conditions in circumferential direction. The far-field signature is obtained by extrapolating the instantaneous pressure on the blade surface.

The RPM method by \citet{ewert_caa_2011} synthesizes the turbulent fluctuations by spatially filtering white noise. The mean turbulent quantities are taken from a preceding RANS simulation. This method is now established and has been successfully applied to model different sources such as jet noise~\cite{ewert_three-parameter_2012}, slat noise~\cite{ewert_broadband_2008}, haystacking~\cite{siefert_sweeping_2009} or airfoil self noise~\cite{cozza_broadband_2012}. For the prediction of fan noise, the method is also applied. \citet{kim_advanced_2015} applied it to investigate the turbulence interaction with a flat plate in two- and three-dimensional space. They generated a von \Karman model spectrum by an optimization technique utilizing a set of Gaussian and Mexican hat filters. The divergence-free turbulence is coupled into a CAA domain by a sponge-layer technique. For centrifugal fans \citet{heo_unsteady_2015} have applied the RPM method to time-periodic flow with cyclostationary turbulence. The synthetic fluctuations are used as sources in an acoustic analogy solved by a boundary element method. A sufficient agreement with measurements is only achieved if cyclostationarity is considered.


To compare the results to experimental data, it is crucial to use realistic model spectra. \citet{dieste_random_2012} have derived complex filter stencils to model von \Karman spectra directly. This turns out to be computationally intensive. A more efficient solution consists in empirically weighting Gaussian filters with different length scales and amplitudes~\cite{kim_advanced_2015,hainaut_caa_2016}. \citet{wohlbrandt_analytical_2016} have derived analytical weighting functions in order to realize typical isotropic turbulence spectra by superposition of Gaussian spectra. They also showed that the reconstruction with five discrete realizations is sufficiently accurate to cover a frequency range with a change of one order of magnitude when the spectra are logarithmically distributed.  The current article will show how this technique can be used to simulate length scales varying in space and time while realizing temporally and spatially constant Gaussian filtered fields.


Broadband noise in the fan stage is caused by the interaction of the turbulence in the rotor wakes with the surfaces at the leading edges of the stator vanes in the presence of a time-periodic mean flow.  
The first objective of the current paper is to utilize the RPM method to develop the cyclostationary stochastic hybrid (CSH) method to include time-periodic turbulence variations and mean flow, which are essential in studying broadband noise generation of rotating parts. 
This method allows for an in-depth study of cyclostationary turbulence and therefore contributes to a greater understanding of broadband noise generation in fans. This is especially important for the development and improvement of analytical tools. The second objective is to separately study the impact of the different effects due to cyclostationarity. A preliminary study utilizing this method was presented by \citet{wohlbrandt_extension_2015}. This article consolidates the method and applies it to another fan, for which measurement data have been made available.


This paper is structured as follows:  The used hybrid approach, the extensions for including cyclostationarity, the general procedure for the setup of such a computation as well as evaluation methods are discussed in Section~\ref{sec:method}.  In Section~\ref{sec:Application}, the CSH method is demonstrated by applying it to the NASA Source Diagnostic Test (SDT) fan.  The effects of cyclostationary wake characteristics on the fan broadband noise are discussed in Section ~\ref{sec:results}.  Additionally, the numerical sound power spectra are compared to experimental data.  Key features of the method as well as significant findings are summarized in Section ~\ref{sec:conclusion}.

\section{Method}
\label{sec:method}

\subsection{Cyclostationarity}
''A cyclostationary signal is a
random signal whose statistical characteristics vary periodically in
time'' \cite{jurdic_investigation_2009}. This is especially relevant in a fan. Although the stochastic signal is different at each revolution of the fan, its characteristics reappear periodically. This is valid for rotor-triggered mean values but also for the turbulent statistics.

By replacing the ensemble average with a cycle average~\cite{gardner_cyclostationarity:_2006}, the subsequently used hybrid method is expanded to reproduce cyclostationary processes. Hence, the mean flow and stationary turbulent characteristics are extended to a periodically changing background flow and cyclostationary turbulence. This method is referred to as cyclostationary stochastic hybrid (CSH) method. Although the changes to the underlying governing equations are small as discussed in subsection~\ref{sec:hybrid}, the resulting level of complexity is very much increased. The combination possibilities are shown in subsection~\ref{sec:coupling}. The influence of periodically changing background flow, variance and length scale is investigated in section~\ref{sec:Application}. 



\subsection{Hybrid approach}
\label{sec:hybrid}
\begin{figure}
%
%
\begin{tikzpicture}[=>latex,
   minimum height   = 1.0cm,
   every node/.style={
       text width=2.1cm
 , align=center},
 font={\sffamily\small}
] 
\tikzstyle{stateEdgePortion} = [black,thick];
\tikzstyle{stateEdge} = [stateEdgePortion,->];
\tikzstyle{edgeLabel} = [pos=0.5, text centered, draw=none, text width=2cm, fill= white, minimum height = 0 cm, font={\sffamily\small}];
\def\l{2.1}
 \node[name = rpm, draw=black]    at (0,0){fRPM};
 \node[name = CAA, draw=black]    at ({2.5*\l},-{3*\l}){CAA \\ PIANO};		
 \node[name = CFD, draw=black]    at (0,-{3*\l}){(U)RANS \\ TRACE};
 
 \node[name = u, text width = 2cm, align =  left, minimum height = 0 cm, draw = none] 
                      at ({1.2*\l},-{2.2*\l}) {Background mean flow};
\node[name = BP] at (-{1.5*\l},-{2*\l}) {Operating point \& geometry};
\node[name = turb]   at (-{1.5*\l},{-1*\l}){Shape of correlation function};
\node[name = schall] at ({4*\l},-{1.5*\l}) {Sound field};

  \draw (rpm.east)
       edge[stateEdge, bend left = 30] 
		node[edgeLabel]{Turbulent fluctuations}
       (CAA.north); 


	\draw (CFD.north)
       edge[stateEdge]
       node[edgeLabel]{Variance, \\ 
                       length scale }
       (rpm.south); 
	\draw (CFD.north east)
       edge[stateEdge]
       (u.south west); 
\draw (u.north west)
       edge[stateEdge, bend right = 15]
       ([xshift=2pt]rpm.south);
\draw (u.south)
       edge[stateEdge, bend left = 15]
       ([yshift=2pt]CAA.west);
\draw (CFD.east)
       edge[stateEdge]
       node[edgeLabel]{Geometry}
       (CAA.west);
\draw (CAA.east)
       edge[stateEdge, bend right = 30]
       (schall.south);
\draw (BP.south)
       edge[stateEdge, bend right = 30]
       (CFD.west);

\draw (turb.north)
       edge[stateEdge, bend left = 30]
       (rpm.west);

 \node[draw,dotted, fit=(schall),inner sep=1mm] (out) {};
 \node[rotate=90,font=\ttfamily,gray,draw=none] at (out.east) {\colorbox{white}{Output}};
\node[draw,dotted, fit=(BP) (turb),inner sep=1mm] (input) {};
\node[rotate=90,font=\ttfamily,gray,draw=none] at (input.west) {\colorbox{white}{Input}};
\end{tikzpicture}
\caption{Overview of the tools involved in the hybrid approach.\label{fig:hybridAnsatz}}
\end{figure}
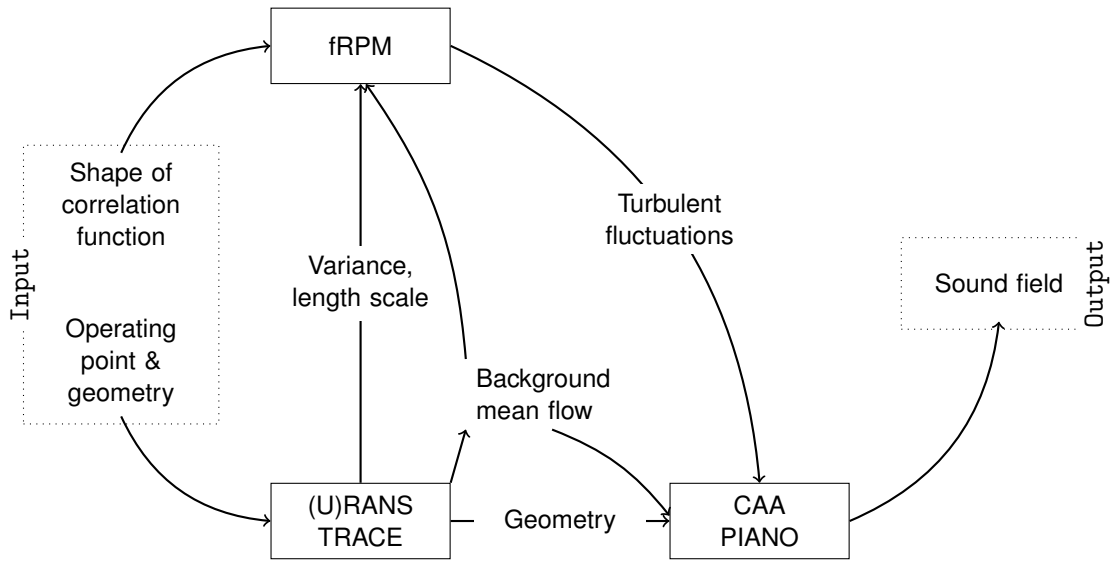

The cyclostationary stochastic hybrid approach, which simulates broadband RSI noise using both stationary and cyclostationary turbulence, is depicted in Fig.~\ref{fig:hybridAnsatz}. It consists of three methods: The (U)RANS method computes the background flow and turbulence statistics. The RPM method synthesizes the turbulence in the time domain. The CAA method convects the synthetic turbulence into the source region and radiates the resulting broadband noise to the sensor positions. 

Solely the geometry and operating condition of the turbomachine and the shape of the correlation function are needed as inputs for this hybrid approach. It outputs broadband time signals at the desired microphone positions, which are converted into sound power levels (PWL) in an equivalent duct.

Next, the separate methods of the hybrid approach and the post-processing are explained.

\subsubsection{(U)RANS: background flow and turbulence statistics}
\label{sec:RANS}
The in-house Computational Fluid Dynamics (CFD) solver TRACE was used~\cite{becker_recent_2010}. The mean flow and mean turbulent statistics were predicted by a Reynolds-Averaged Navier Stokes (RANS) simulation, while the periodic flow, periodic turbulent statistics and the tonal noise were predicted by an unsteady RANS (URANS) simulation. 

In this investigation, the (U)RANS calculations were performed on a quasi-3D (q3D) domain. A q3D computational domain consists of a few cells in the radial direction and its radial boundaries follow streamlines of a preliminary 3D RANS simulation. In most cases, the aerodynamic results of such a q3D computation closely resemble the 3D results at the same position~\cite{holewa_impact_2014}.
 

\subsubsection{Random Particle Mesh (RPM) method: synthesized turbulence}
The Random Particle Mesh method~\cite{ewert_caa_2011} allows to synthetically realize the time-space-dependent turbulent fluctuations based on the background flow, the local variance, and the local length scales provided by the (U)RANS simulation. In general, the RPM method is able to generate anisotropic, evolving synthetic turbulence of local integral turbulence length scale $\Lambda$ 
realizing arbitrary model spectra. The turbulence is generated by spatially filtering a random stochastic field with a Gaussian filter of the aforementioned length scale. The turbulence is scaled with the local variance, which corresponds to the turbulent kinetic energy $k_t$ for this particular application, and convects with the local background velocity $\u_0$.

The derivation of the RPM method is neither restricted to a spatially constant variance or length scale nor to a constant background mean flow. The statement of separating velocity $\u = \u_0 + \u'$ into a background flow $\u_0$ and a fluctuating part $\u'$ is still valid if the background flow $\u_0$ is also fluctuating in time. In this manner, $\u'$ can be understood as the fluctuation in a moving frame of reference. The only requirement is that $\u_0$ has to be known in advance. Therefore, $\u_0$ cannot be influenced by the fluctuating velocity components $\u'$. 

A fast Random Particle Mesh (fRPM) method was implemented by \citet{siefert_sweeping_2009} utilizing recursive filters on a Cartesian grid to speed up the computation. This implementation was used in this investigation to synthesize the turbulent field.

The method has been extended to arbitrary spectra by \citet{wohlbrandt_analytical_2016} by deriving analytical weighting functions $f(l,\Lambda)$. Thus, a von K\'arm\'an spectrum with an integral length scale $\Lambda$ can be realized by superposing $N_i$ Gauss spectra of various length scales $l_i$:
\begin{align}
E(k,\Lambda) &\approx \sum\limits_{i=1}^{N_i} f(l_i,\Lambda) E_G(k,l_i) \Delta l. \label{eq:weighting_disketerAnsatz}
\end{align}
By using spatially and temporally constant length scales $l_i$, it is possible to use the recursive Young-van-Vliet Filter \cite{young_recursive_1995}, which is restricted to constant length scales but beneficial in terms of computational time and robustness. 
The realization of spatially and temporally varying length scales $\Lambda(\x,t)$ with this filter necessitates appropriate weighting functions:
\begin{align}
E(k,\Lambda(\x,t)) &\approx \sum\limits_{i=1}^{N_i} f(l_i,\Lambda(\x,t)) E_G(k,l_i) \Delta l. \label{eq:weighting_timeDep}
\end{align}
This approach allows for arbitrarily high spatial and temporal gradients in the integral length scale. 
\subsubsection{Computational Aeroacoustics (CAA): convection of turbulent disturbances, generation and propagation of acoustic waves }
The turbulent and acoustic fluctuations are resolved in the time domain by the CAA solver PIANO~\cite{delfs_numerical_2008}. In this study, it was used to solve the linearized Euler equations. The well-known dispersion-relation-preserving finite difference scheme by \citet{tam_dispersion-relation-preserving_1993} was utilized for the spatial discretization and the low-dispersion low-dissipation Runge-Kutta method~\cite{hu_low-dissipation_1996}, for time-integration. 
The LEE for time-varying Navier-Stokes background flow can be derived by neglecting the turbulent stresses, viscous effects, heat fluxes and non-linear terms in the non-linear Navier-Stokes equations in perturbation form~\cite{ewert_linear-_2014} without limitations to slow changes of $\u_0$. Using the tensor notation $\u_0 = u^0_i \e_i$, it reads:
\begin{subequations}
\begin{align}
\pAbl{\rho'}{t}  + \pAbl{}{x_i}\left(\rho'u^0_i+\rho_0 u_i'\right) &= 0\\
\pAbl{u_i'}{t}+u^0_j\pAbl{u_i'}{x_j}+u_j'\pAbl{u^0_i}{x_j}+\frac{1}{\rho_0}\left(\pAbl{p'}{x_i}-\frac{\rho'}{\rho_0}\pAbl{p_0}{x_i}\right)&=0\\
\pAbl{p'}{t}+u^0_i\pAbl{p'}{x_i}+ u_i'\pAbl{p_0}{x_i}+\gamma\left(p'\pAbl{u^0_i}{x_i}+p_0\pAbl{u_i'}{x_i}\right)&=0.
\end{align}
\end{subequations}
It can be taken advantage of the periodicity in the fan by setting all flow field variables, represented by $\phi_0(t)$, to 
\begin{align}\label{eq:fourierCoeff4CAA}
\phi_0(t) = \sum\limits_{k = 0}^{\infty}\phi^0_{k} e^{-i k \omega t},
\end{align}
where $\phi^0_{k}$ is the $k^\text{th}$ harmonic of the base frequency $\omega$ and $k = 0$ represents the steady part $\phi^0_0 = \overline{ \phi_0(t)}$. That way, only the complex Fourier coefficients need to be stored to reconstruct the flow at each time step.

\subsection{Coupling of fRPM and CAA methods}
\label{sec:coupling}
Fluctuating vorticity as generated by the fRPM method over a given background flow is coupled with the CAA solver PIANO by applying the LEE-relaxation formulation~\cite{ewert_linear-_2014}. This method adds a relaxation term to the impulse equations,
\begin{align}
\pAbl{u'_i}{t} + \cdot\cdot\cdot = \epsilon_{ijk}\pAbl{}{x_j}\left[\sigma\left(\Omega'_k -\Omega^\text{ref}_k\right)\right],
\end{align}
where $\sigma$ is the forcing parameter, $\Omega'_i$ is the vorticity given by the left hand side of this equation as
\begin{align}
\Omega'_i = \epsilon_{ijk}\pAbl{u'_k}{x_j},
\end{align}
and $\Omega^\text{ref}_i$ is the externally imposed fluctuating reference vorticity. This reference vorticity is determined by the fRPM method by applying the so called 'source A' model~\cite{ewert_caa_2011}. In short, this means that a stream function is modeled to realize the divergence-free turbulent velocity fluctuations, from which the vorticity is derived.

In this paper, the turbulence is coupled into the CAA domain in a region upstream of the blades. Therefore, only frozen Taylor vortices but no de-correlation effects or local turbulent characteristics can be modeled. To enable a coupling with negligible energy loss, the forcing parameter $\sigma$ must be as high as the stability limit permits. This stability limit behaves similarly as that for viscous equations as discussed by \cite{moghadam_implementation_2012}. In fact, the forcing parameter has the unit of viscosity. The approach during this study was to iteratively increase the forcing parameter, while keeping the simulation stable without changing the initial linear stability time step given by the CFL number.   
 
There are several advantages of using a LEE-relaxation formulation rather than a modified \citet{tam_radiation_1996} radiation boundary condition (BC) or an additional sponge zone.  While the results of all methods are identical, the LEE-relaxation formulation is invisible to acoustic pressure waves and can be placed anywhere inside the computational domain.  Thus, a smaller region of the grid upstream of the blade has to resolve small turbulent structures, which requires less computational effort. Lastly, it can locally and noiselessly cancel out vorticity waves by setting $\bfm \Omega^\text{ref} = 0$ where needed.
 
\subsection{Types of cyclostationarity realized by the CSH method}\label{sec:config}
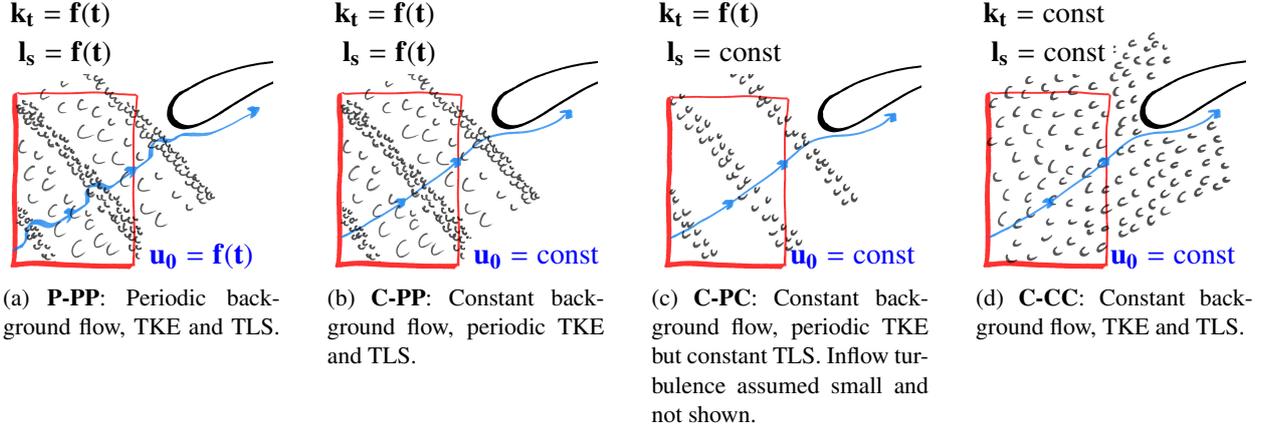
\begin{figure}
\subfigure[\textbf{P-PP}: Periodic background flow, TKE and TLS. \label{fig:configPPP}]{
\begin{overpic}[width=0.21\textwidth]{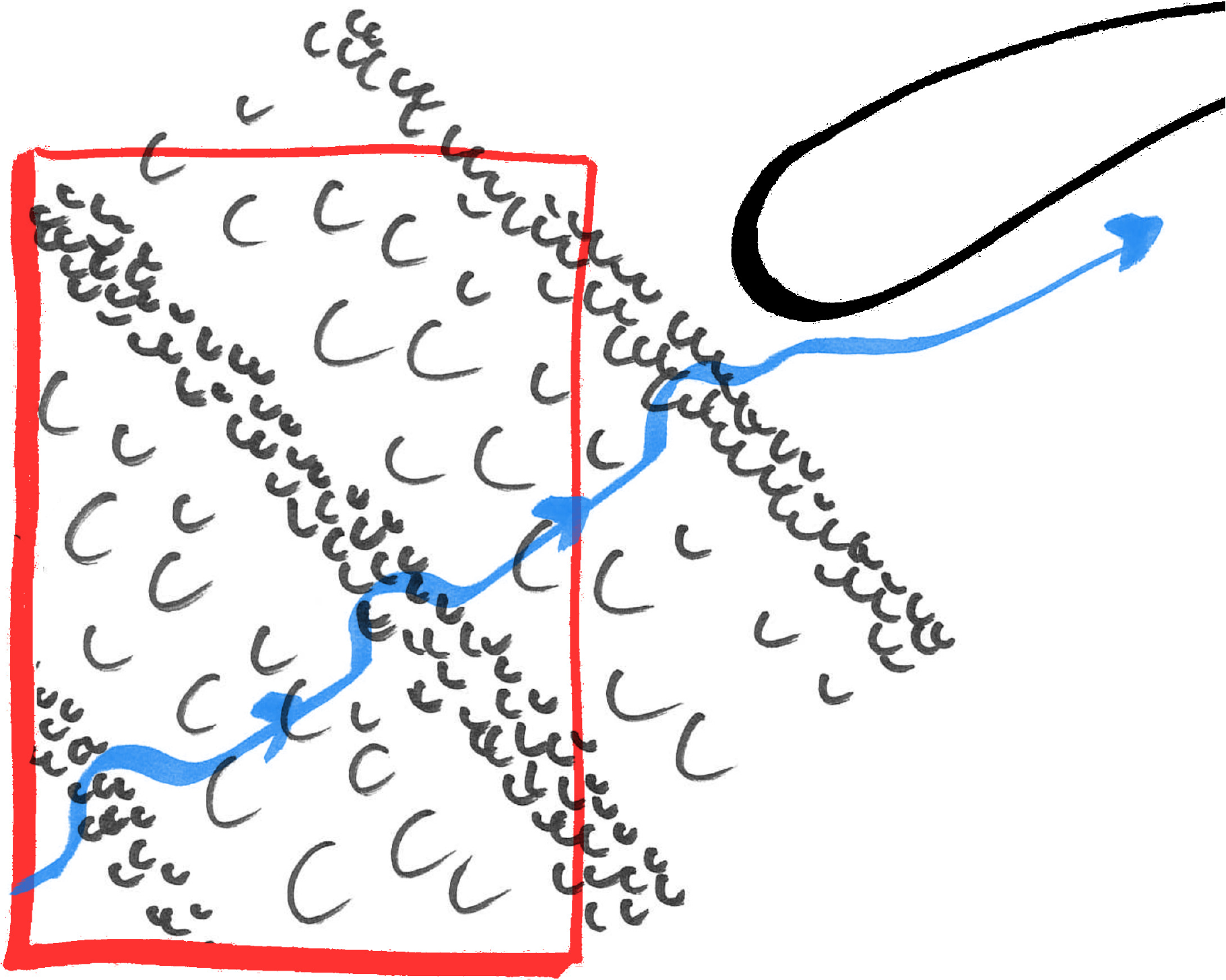}
\put(0,80){\colorbox{white}{$\mathbf{ l_s = f(t)}$}}
\put(0,95){$\mathbf{ k_t = f(t)}$}
\put(50,2){\colorbox{white}{\textcolor{blue}{$\mathbf{ u_0 = f(t)}$}}}
\end{overpic}
}\hfill
\subfigure[\textbf{C-PP}: Constant background flow, periodic TKE and TLS. \label{fig:configCPP}]{
\begin{overpic}[width=0.21\textwidth]{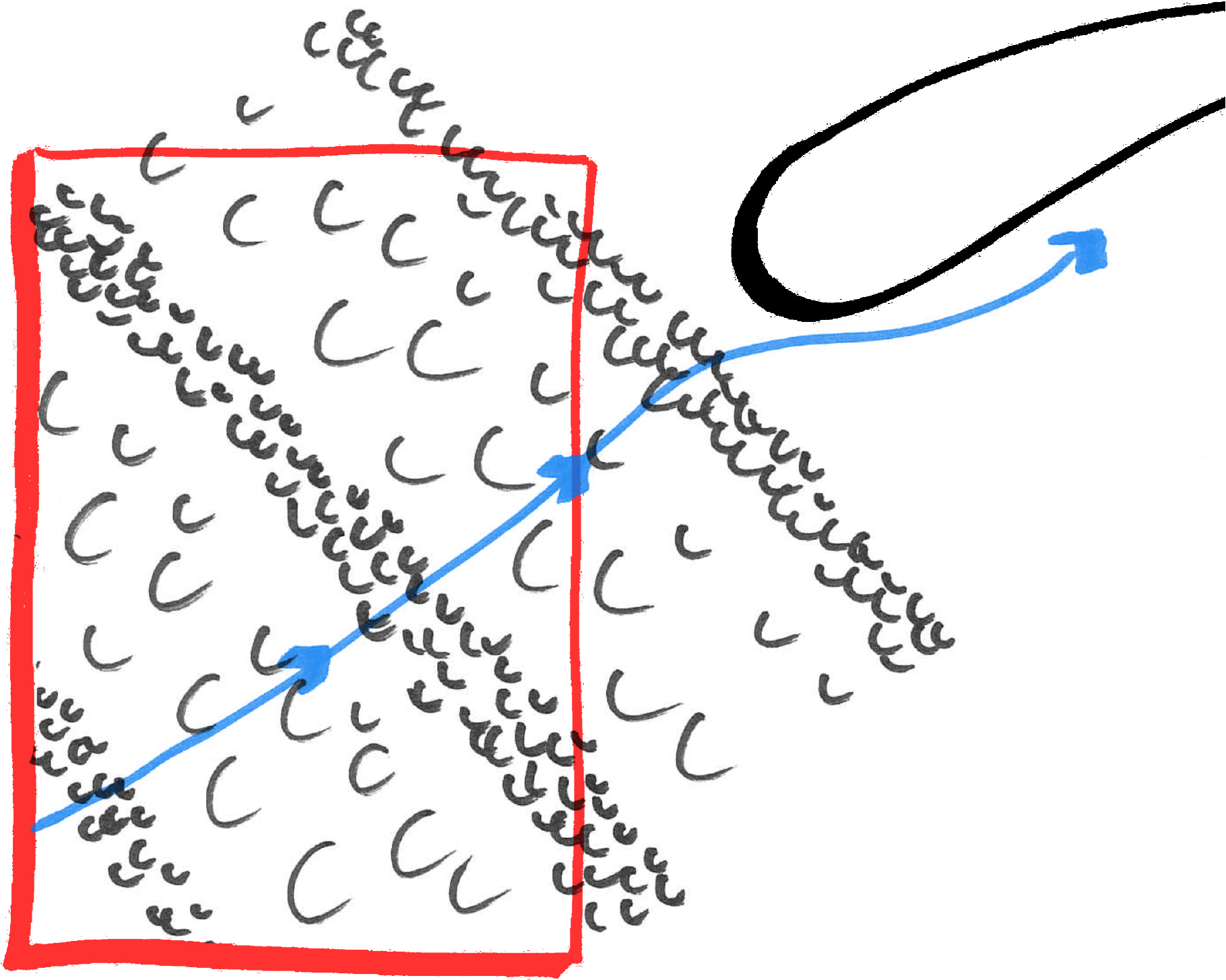}
\put(0,80){\colorbox{white}{$\mathbf{ l_s = f(t)}$}}
\put(0,95){$\mathbf{ k_t = f(t)}$}
\put(50,2){\colorbox{white}{\textcolor{blue}{$\mathbf{ u_0 = \const}$}}}
\end{overpic}
}\hfill
\subfigure[\textbf{C-PC}: Constant background flow, periodic TKE but constant TLS. Inflow turbulence assumed small and not shown.  \label{fig:configCPC}]{
\begin{overpic}[width=0.21\textwidth]{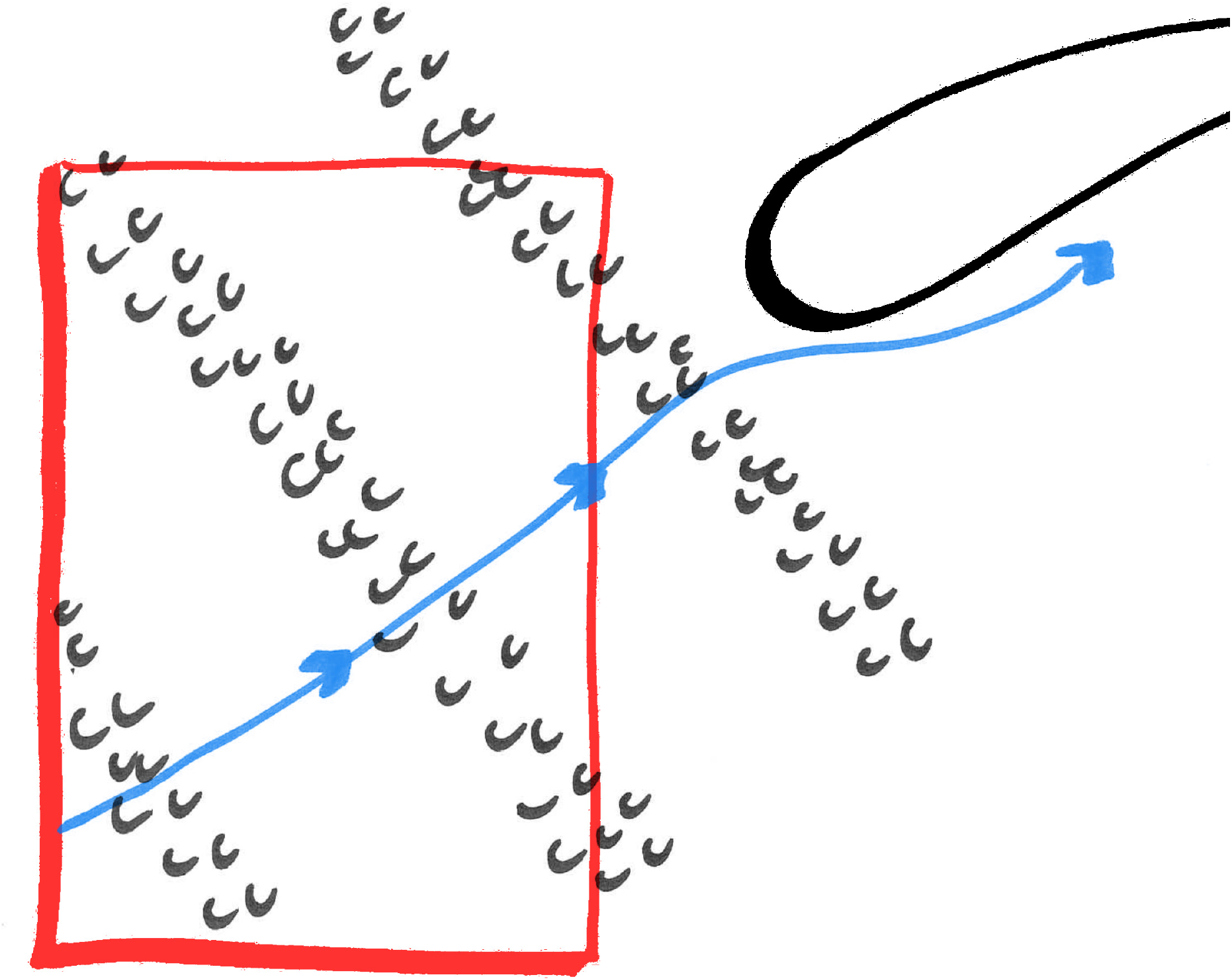}
\put(0,80){\colorbox{white}{$\mathbf{ l_s = \const}$}}
\put(0,95){$\mathbf{ k_t = f(t)}$}
\put(50,2){\textcolor{blue}{$\mathbf{ u_0 = \const}$}}
\end{overpic}
}\hfill
\subfigure[\textbf{C-CC}: Constant background flow, TKE and TLS.\label{fig:configCC}]{
\begin{overpic}[width=0.21\textwidth]{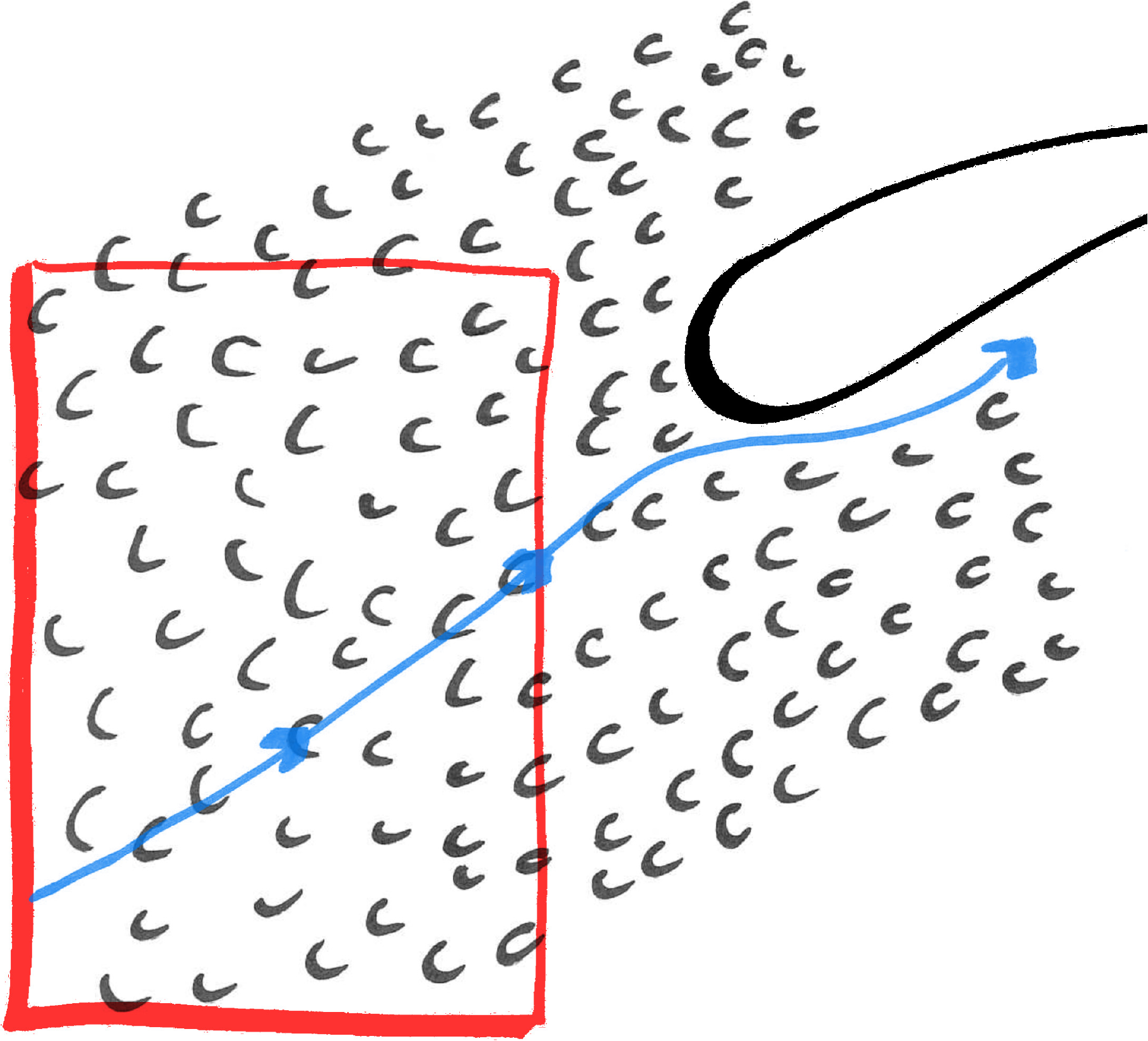}
\put(0,80){\colorbox{white}{$\mathbf{ l_s = \const}$}}
\put(0,95){$\mathbf{ k_t = \const}$}
\put(48.5,2){\textcolor{blue}{$\mathbf{ u_0 = \const}$}}
\end{overpic}
}
\caption{Varying complexity for coupling cyclostationary turbulence into the CAA domain. The red box depicts the fRPM domain and the blue line indicates the flow path. The inflow turbulence between the rotor wakes is only schematically shown in \subref{fig:configPPP} and \subref{fig:configCPP} to emphasize the difference due to the TLS. \label{fig:Configurations}}
\end{figure}

The possible cases offered by the CSH method are discussed in this subsection. A summary and the hereafter used abbreviations are listed in Table~\ref{Tab:Configurations} and Figure~\ref{fig:Configurations}. The investigated configurations are detailed in Section~\ref{sec:confRealised}.

\begin{table}[h]
\caption{Four simulation configurations aimed at investigating the impact of cyclostationarity in RSI broadband noise.
\label{Tab:Configurations} }
\begin{center}
\begin{tabular}{|c|c|c|c|}
\hline
configurations & background mean flow      & \multicolumn{2}{c|}{turbulence}   \\
\cline{3-4}
               &                & TKE           & TLS      \\
\hline
\hline
{\bf P-PP}            & time-{\bf P}eriodic  & \multicolumn{2}{c|}{time-{\bf P}eriodic}  \\
\hline
{\bf C-PP}            & {\bf C}onstant       & \multicolumn{2}{c|}{time-{\bf P}eriodic}  \\
\hline
{\bf C-PC}           & {\bf C}onstant        & time-{\bf P}eriodic & {\bf C}onstant             \\
\hline
{\bf C-CC}            & {\bf C}onstant       &  \multicolumn{2}{c|}{{\bf C}onstant}   \\
\hline
\end{tabular}
\end{center}
\end{table}

For the sake of clarity and comparability, all subsequent cases realize the sources in the stator frame of reference using a patch at the same position and of the same size. There is also the possibility to model the sources in the rotor frame of reference with a so-called moving patch. This has been discussed in \cite{wohlbrandt_extension_2015} and is not brought up again as it does not give any additional physical insight.

\paragraph{{\bf P-PP}: Time-periodic mean flow, TKE and TLS,} 
see Fig.~\ref{fig:configPPP}. 
This variant most closely resembles reality since it fully accounts for the periodicity in the background flow and turbulence statistics as prescribed by the URANS. As in Eq.~\ref{eq:fourierCoeff4CAA} the Fourier coefficients of the harmonics are used to reproduce the flow and statistics at each time step. This accounts for the mean wake velocity deficit, the periodicity of the turbulent kinetic energy statistics, and the variation of the integral length scale across the blade passage. The turbulence generation in fRPM and the background mean flow in fRPM and CAA domains are synchronized. It could mistakenly be assumed that this configuration generates the tones produced by the mean wake deficit impingement on the stator, even without the stochastic sources. However, this is not the case. The pure tones are part of the URANS simulation and appear as part of the time-varying background flow in the CAA domain. But since the CAA realizes the fluctuations on top of the background flow, there are no tones in the CAA simulation itself. Only the superposition of CAA and URANS data contains these tones. 

\paragraph{{\bf C-PP}: Constant mean flow with time-periodic TKE and TLS,}
see Fig.~\ref{fig:configCPP}.
The complexity can be reduced by neglecting the periodic background flow in both the fRPM patch and the CAA domain. This is accomplished by using only the $0^\text{th}$ harmonic of the flow, i.e. the steady part. Compared to the P-PP case, the influence of the periodic background flow can be studied. To guarantee consistency in the input data for all considered cases, the TKE and TLS are extracted directly from the URANS simulation. 
Since a TLS cannot be expressed in terms of Fourier coefficients, the Fourier coefficients of the TKE and the specific dissipation rate $\omega_t$ are used instead.  Then, the TLS can be computed for each time step as follows:
\begin{equation}\label{eq:Lambda}
\Lambda = \frac{C_\text{Re}}{C_\mu}\frac{\sqrt{k_t}}{\omega_t}.
\end{equation}
A model parameter $C_\mu=0.09$ and a Reynolds-number dependent scaling parameter $C_\text{Re} \approx 0.4$~\cite{pope_turbulent_2000} were used.
Note that for this case and subsequently discussed cases, it would be sufficient to use a RANS simulation. In this respect, the Fourier coefficients result from the transformation of the stationary wake in the rotor frame of reference to the stator frame of reference. 

\paragraph{{\bf C-PC}: Constant mean flow with time-periodic TKE but uniform constant TLS,}
see Fig.~\ref{fig:configCPC}.
By additionally averaging the time-periodic TLS, e.g. by replacing it by its $0^\text{th}$ harmonic, only the periodicity of the turbulent kinetic energy statistics remains. \citet{dieste_random_2012} have investigated this case analytically.

\paragraph{{\bf C-CC}: Constant mean flow with uniform constant TKE and TLS,} see Fig.~\ref{fig:configCC}.
This configuration corresponds to homogeneous stationary turbulence impinging on a stator blade. This approach uses the datum fRPM method without cyclostationarity, which has been validated before~\cite{wohlbrandt_simultaneous_2013}. It is very similar to the approach used by the authors for predicting the FC1 benchmark case presented at the fan broadband noise workshop of AIAA Aviation 2014.  The CFD solution could come from a steady-state RANS calculation, but for consistency the $0^\text{th}$ harmonic of a URANS is used here.

%
%

\section{Application}\label{sec:Application}

The CSH method described in the previous section was applied to NASA's 22-in Source Diagnostic Test (SDT) fan at approach condition.  The effects of cyclostationary parameters were examined and the numerical data were compared to experimental data presented at the Fan Broadband Noise Prediction Workshop organized in the framework of the AIAA 2014 and 2015 Aeroacoustics Conferences.  The experimental setup of the Realistic Test Case 2 (RC2) was described in detail by \citet{nallasamy_computation_2005}.  The operating conditions are given in table~\ref{tab:operatingPoint} and the input specifications, in the workshop problem statement~\cite{envia_panel_2015}.    


\begin{table}
		\centering
		\caption[Fan characteristics]{Fan characteristics and operating points of the NASA-SDT fan. Design point is not used, but shown here for reference. \label{tab:operatingPoint}}
		\begin{tabular}{|l|r|r|}
		\hline
		Fan diameter   		& \multicolumn{2}{ c|}{\SI{0.56}{m}}	\\
		Rotor blade count $N_B$
							& \multicolumn{2}{ c|}{22} 			\\
		Stator vane count $N_V$ 
							& \multicolumn{2}{ c|}{54} 			\\
		\hline
		\hline
			&\textit{Design} &\textit{Approach} \\
		\hline
		Fan-Pressure ratio $\Pi$
							& 1.48			& 1.15		\\
		Axial Mach number $M_x$ in rotor plane
							& 0.59			& 	0.31	\\
		Relativ tip Mach number $M_\text{tip,rel}$
							&	1.39		&   0.80		\\ 
		\hline
		\end{tabular}
		\end{table}

\subsection{Description of applied procedure}

The CAA simulations were performed on two-dimensional cascade mesh at midspan of the annular duct at the stator leading edge. It is computed in two-dimensional space to reduce simulation cost and to, therefore, allow for parameter variations. 
The solidity of the vanes is greater than one. Thus the cascade effect cannot be neglected~\cite{blandeau_comparison_2011} and all stator vanes had to be considered to correctly predict the sound propagation.  

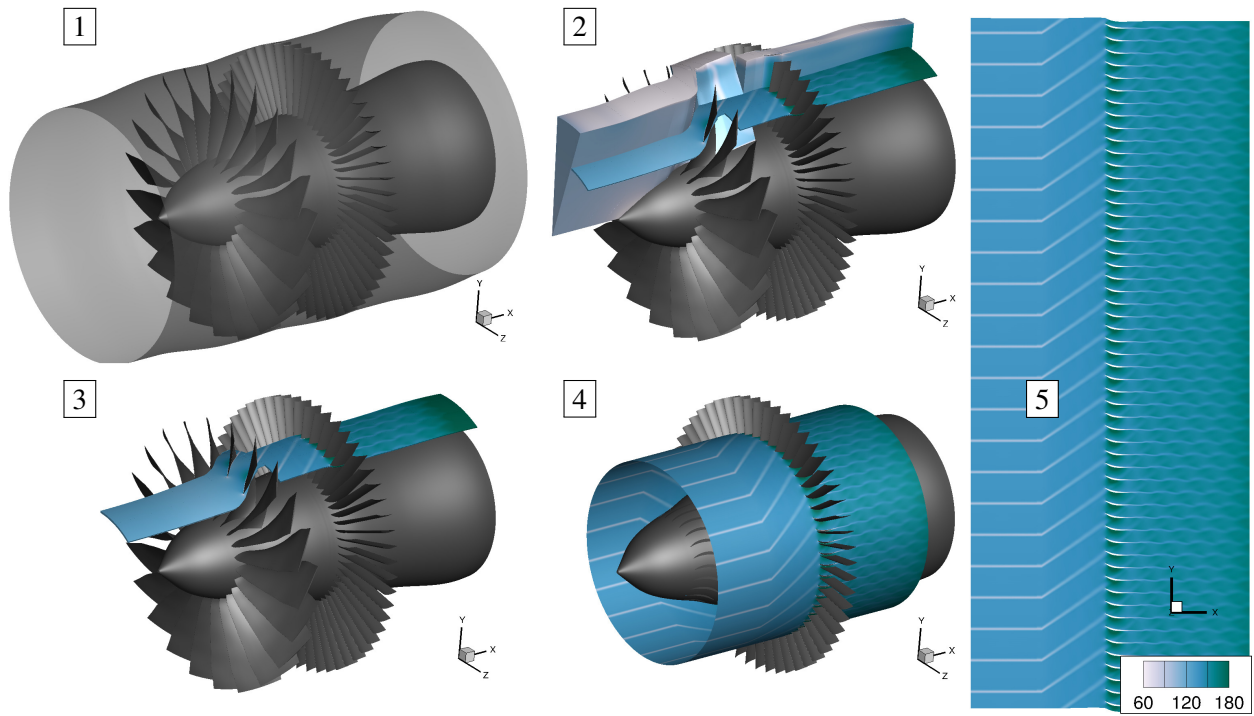
\begin{figure}[h]
	\centering
	\begin{overpic}[width=\textwidth]{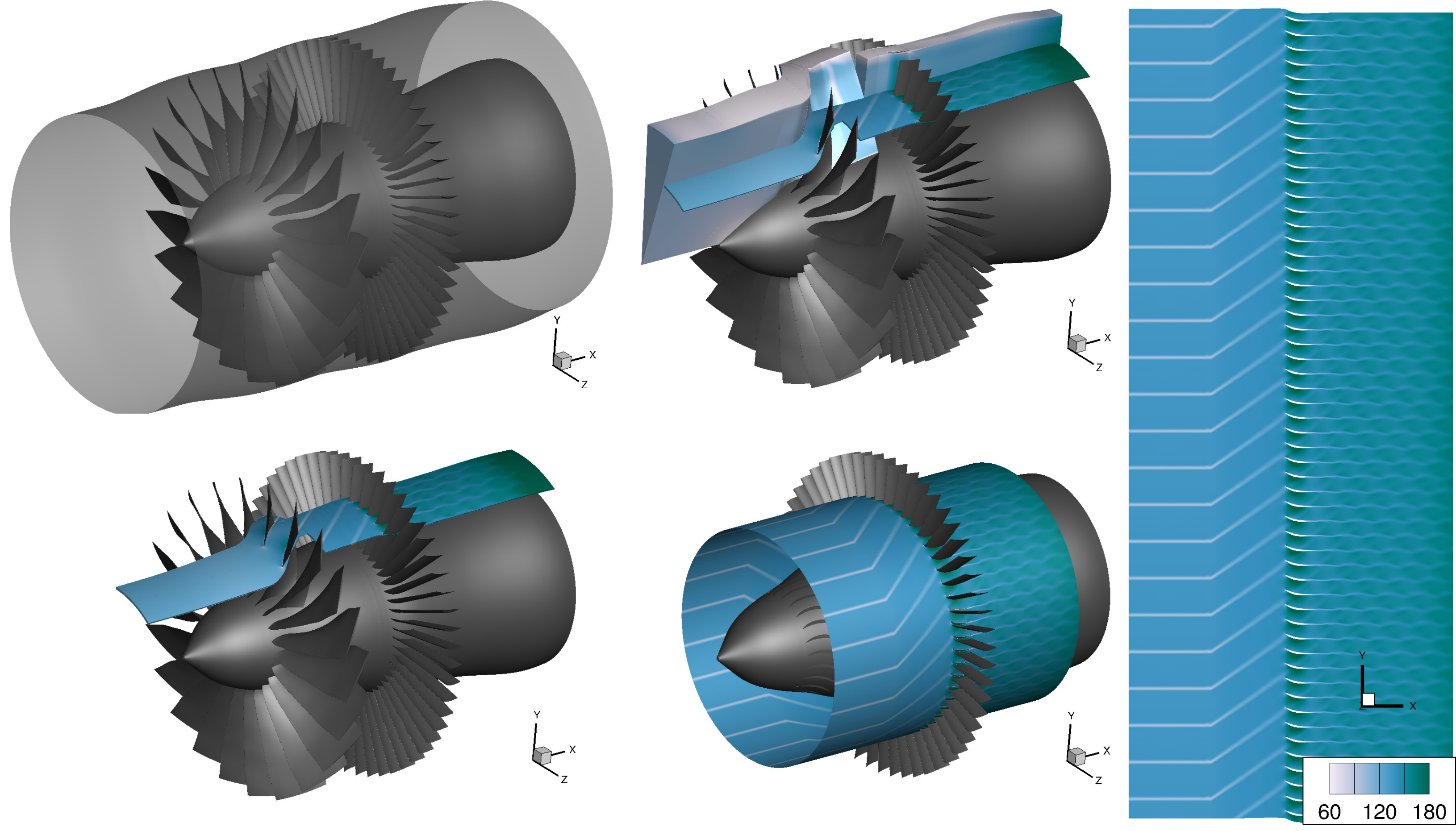}
	\put( 5,55){\fcolorbox{black}{white}{1}}
	\put(45,55){\fcolorbox{black}{white}{2}}
	\put( 5,25){\fcolorbox{black}{white}{3}}
	\put(45,25){\fcolorbox{black}{white}{4}}
	\put(82,25){\fcolorbox{black}{white}{5}}
	\end{overpic}
	\caption{Transfomation of the fan geometry (1) into a two-dimensional cascade. Instantaneous axial component of the background flow in [m/s] is shown.\label{fig:RC2_transform}}
\end{figure}


To prepare the two-dimensional cascade domain for the CAA, a series of steps was followed as illustrated in Figure~\ref{fig:RC2_transform}. The numbers in the subsequent description correspond to the numbers in the figure. Text that appears in italics refers to topics which are further explained below the enumeration.
\begin{enumerate}
\item Firstly, the CAD geometry of the SDT fan with the baseline stator configuration has to be obtained.  For this case, the number of stator vanes was increased from 54 to 55.  The new number of vanes allowed for a reduced computation with fully periodic boundary conditions using only two rotor blades and five stator vanes.  This made the URANS and CAA computations more efficient.  The increase in the number of stator vanes was not expected to significantly change the broadband noise characteristics of the fan stage.       

\item Secondly, the CAD geometry was used to set up a three-dimensional RANS simulation for one blade passage. 

\item A q3D domain was generated by extracting streamlines at 49\%, 50\%, and 51\% of the duct height at the stator leading edge from the RANS computation.  From that a q3D URANS computation was set up with two cells in the radial direction for two rotor and five stator passages to achieve circumferential periodicity. 

\item In the next step, the streamline at \textit{50\% of the duct radius} was extracted from the q3D URANS in the stator domain only and expanded to either the full annular duct or to a fraction of the annular duct. The \textit{rotor was ignored}. The Fourier coefficients of flow and turbulent variables were then interpolated from this extracted and expanded CFD mesh onto the CAA mesh and the fRPM patch.  In areas where the CAA mesh lay outside of the  CFD stator domain, an \textit{extrapolation} was applied.  

\item Lastly, a transformation into \textit{stream-surface coordinates} was applied to the flow and geometry to produce a 2D cascade.               
\end{enumerate}
Now the keywords in italics are detailed.

\begin{figure}
		\subfigure[Turbulent kinetic energy (TKE).\label{fig:RC2_hotwire_TKE}]{
			\includegraphics[width=0.45\textwidth]{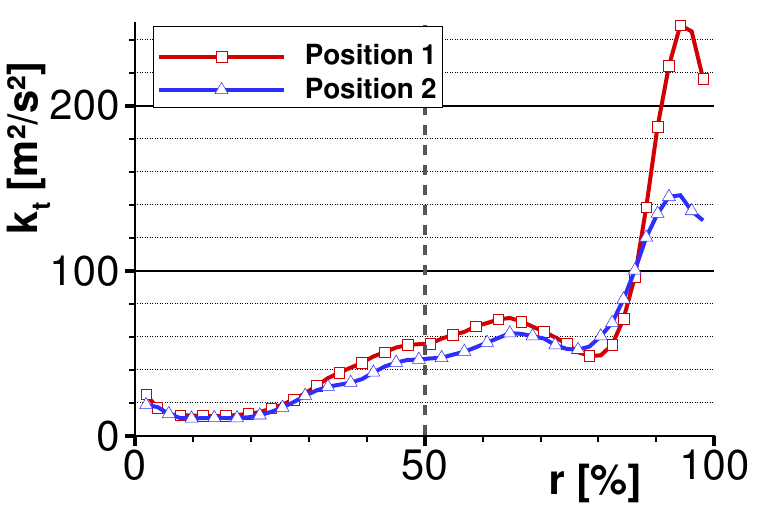}}
		\subfigure[Turbulent integral lengthscale (TLS).\label{fig:RC2_hotwire_TLS}]{
			\includegraphics[width=0.45\textwidth]{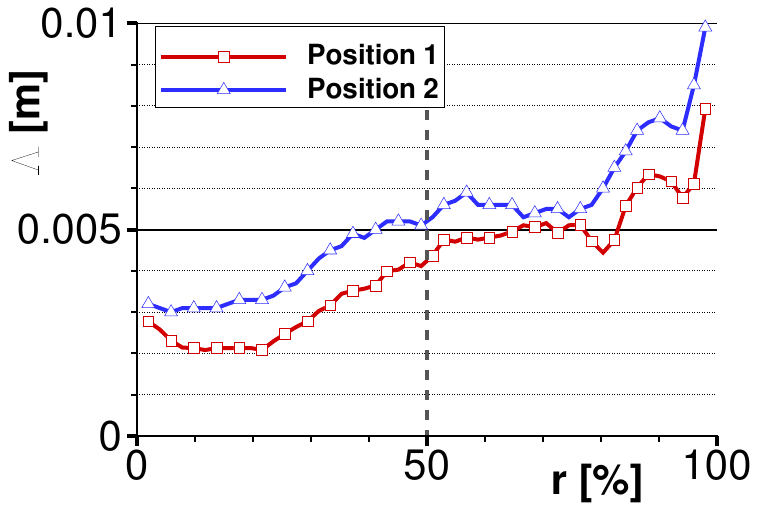}}
	\caption{Turbulent characteristics determined through hotwire measurements by \citet{podboy_steady_2003} for the benchmark testcase~\cite{envia_panel_2015} at two axial positions: (1) half-way between rotor and stator and (2) in front of the stator LE. The dashed vertical line indicates the position of CAA simulation 50\%. (Reproduced with permission) \label{fig:hotwire}}
\end{figure}

\begin{description}
\item{Using a streamline at \textit{50\% of the duct radius}}
	for the q3D simulation is considered representative for the whole duct. For fan acoustic power measurements with omnidirectional microphones, this position is representative according to ISO 5136:1990~\cite{barret_noise_1960,arnold_experimentelle_1999}. Repeating the simulation at differing duct heights would increase the accuracy. The hotwire measurements reproduced in Fig.~\ref{fig:hotwire} show small variations of the turbulent characteristics in a considerably large area.  The maximum at the outer rim is due to a flow detachment in the rotor tip region. This effect cannot be accounted for here.
\item{\textit{The rotor was ignored}}
	as its consideration is computationally expensive. The error in sound radiation must be kept in mind but seems acceptable for subsonic flow without shocks \cite{moreau_impact_2016}. The main effect of the rotor is to block the acoustic waves propagating in the upstream direction. This shielding effect increases with the relative  Mach number of the rotor.    
\item{\textit{The Extrapolation}}
	of the flow quantities is used to allow for the use of a larger CAA domain, especially for the damping zones at the inflow and outflow boundaries. A nearest-neighbor interpolation was used. For the complex coefficients of the non-stationary flow data of the P-PP configurations, the phase cannot be considered as can be seen in the time-reconstructed flow at Steps 4 and 5 of Fig.~\ref{fig:RC2_transform}. The solution is nevertheless continuous and does not have a noticeable influence on the acoustic radiation.
\item{A \textit{stream-surface coordinate}}
	transformation from the q3D-grid into 2D-grid coordinates $m'$ and $\vartheta$ was applied as described in the Appendix~\ref{app:mtheta}. 
	 The benefits are that the flow quantities are transformational invariants and the circumferential distance is independent of the axially changing duct radius of the streamline. The latter allows for the application of periodic boundary conditions. For convenience, we use for the 2D grid coordinates
	\begin{align}
	x &= R_\text{LE} m', &&& y &= R_\text{LE} \vartheta,
	\end{align}
	with the radius at the leading edge $R_\text{LE}$. Hence, the chord length and pitch are invariant to the transformation.
\end{description}

\subsection{Definition of test matrix}
\label{sec:confRealised}
Eight different configurations were realized.  An overview of those configurations is given in Table~\ref{tab:conf}.  The abbreviations used to denote the test cases were introduced in Section~\ref{sec:config}.  The shown test matrix consists of five primary test cases investigating the effect of cyclostationarity and three secondary test cases confirming supplementary aspects. The same CAA grid resolution was used for all test cases.  

\begin{table}
\caption{Test matrix of simulated configurations \label{tab:conf}}
\centering\setlength\extrarowheight{2pt}
\begin{tabular}{|c|l|l|m{1cm}|m{1cm}|m{1cm}|m{1cm}|}
\hline
& type 			& specifications &
\rotatebox[origin=b]{90}{number} 
\rotatebox[origin=b]{90}{of vanes} &	
\rotatebox[origin=b]{90}{periodic}
\rotatebox[origin=b]{90}{mean flow}	& 
\rotatebox[origin=b]{90}{periodic}
\rotatebox[origin=b]{90}{TKE}          & 
\rotatebox[origin=b]{90}{periodic}
\rotatebox[origin=b]{90}{TLS} \\ \hline\hline
\multirow{5}{*}{\parbox[c]{2mm}{\rotatebox[origin=c]{90}{primary}}}
	& P-PP	&	        & 5	& 		yes		&	yes	&	yes \\ \cline{2-7}
	& C-PP	&	        & 5	& 		no		&	yes	&	yes \\ \cline{2-7}
	& C-PC	&	        & 5	& 		no		&	yes	&	no \\ \cline{2-7}
	& C-CC	&	        & 5	& 		no		&	no	&	no \\ \cline{2-7}
	& C-CC-tls	& alternative TLS	        & 5	& 		no		&	no	&	no \\ \hline \hline
\multirow{3}{*}{\parbox[c]{2mm}{\rotatebox[origin=c]{90}{secondary}}}
	& C-CC-55 & full cascade
						& 55	& 		no		&	no	&	no \\ \cline{2-7}
	& C-CC-red  & patch with reduced damping area	
						& 5	& 		no &			no	&	no \\ \cline{2-7}
	& C-CC-double  & double patch		        
						& 5	& 		no		&	no	&	no \\ \hline
\end{tabular}
\end{table}

\begin{figure}
		\subfigure[P-PP test case\label{fig:PPPinCAA}]{
			\includegraphics[width=0.50\textwidth,trim=4 4 4 4,clip]{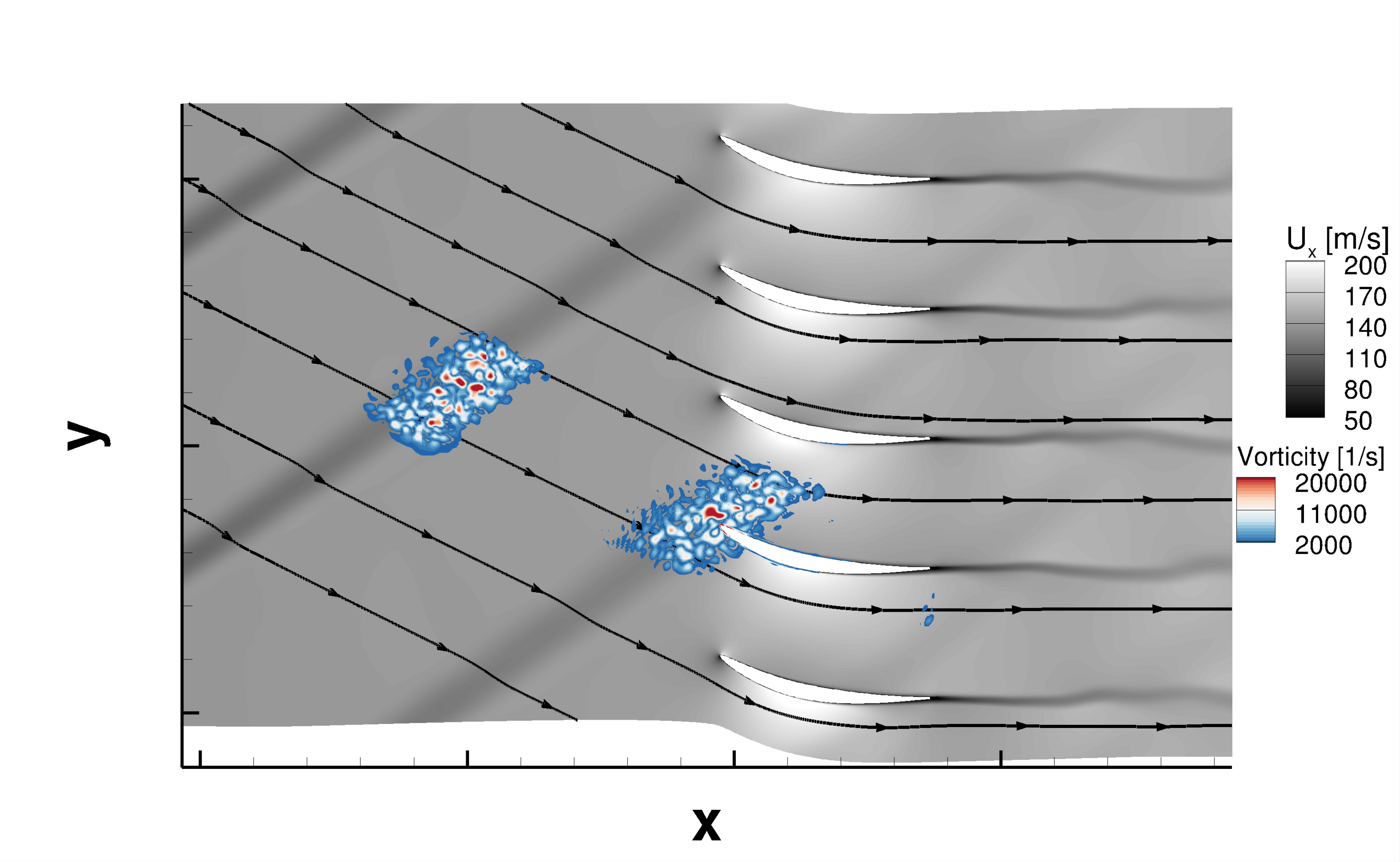}}
		\subfigure[C-CC test case\label{fig:CCCinCAA}]{
			\includegraphics[width=0.50\textwidth,trim=4 4 4 4,clip]{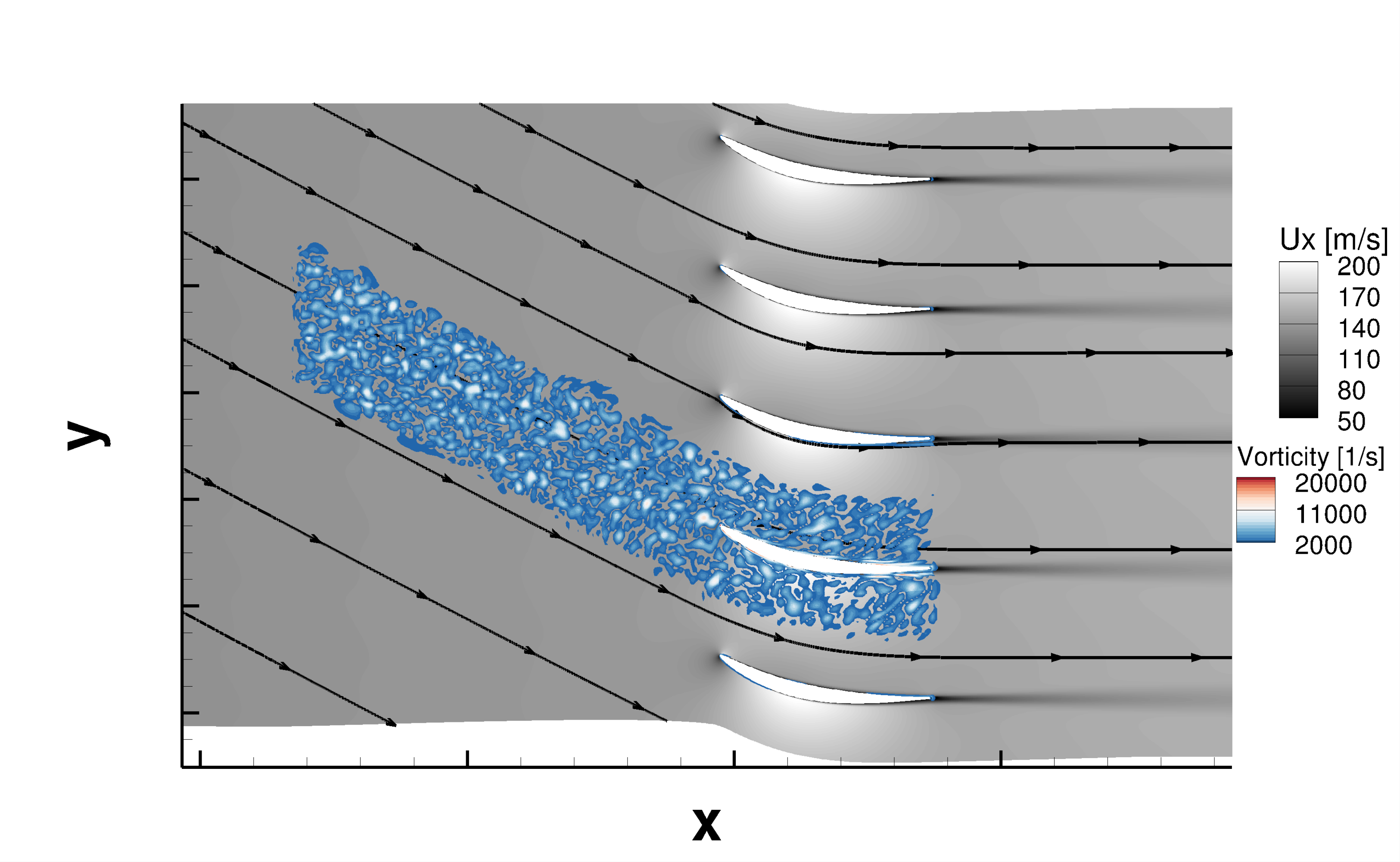}}
	\caption{CAA domain of P-PP and C-CC test cases.  Contours show background axial velocity and turbulence vorticity magnitude.\label{fig:PPPCCCinCAA}}
\end{figure}

The five primary test cases were simulated with only five stator vanes in order to reduce simulation times.  The boundary conditions for the reduced number of blades were still fully periodic, since the number of total vanes was increased to 55. For the primary test cases, the fRPM patch remained unchanged so that the influence of every parameter could be investigated.  The cases have been described in section~\ref{sec:config}. Figures~\ref{fig:PPPinCAA} and \subref{fig:CCCinCAA} show the instantaneous vorticity fields for the P-PP and the C-CC cases, respectively. The P-PP case is the most realistic test case realizing full cyclostationarity. The structure of the wake is reproduced in the mean flow, as shown by the contour of the axial velocity, and in the wake turbulence, as shown by the contour of the vorticity magnitude. The primary C-CC case uses a constant TKE, a constant TLS in fRPM domain, the so-called patch, and a constant mean flow in the CAA and fRPM domains.  Figure~\ref{fig:CCCinCAA} shows that the rotor wake structures are neglected and averages are used.  The C-CC-tls case also uses constant turbulent and flow variables but the TLS was determined using a different technique, which will be discussed in detail in Section~\ref{sec:results}.

The three secondary test cases were done in order to substantiate the findings of the four primary test cases.  The C-CC-55 simulation was done using a full cascade with 55 vanes.  Furthermore, two test cases were simulated with modified patches and constant cyclostationary characteristics.  The seeding area of the patches remained unchanged.  Details regarding the fRPM patch generation can be found in subsection~\ref{subsec:SetupFRPMCAA}.
For case C-CC-red, the initial patch was modified by significantly reducing the safety margins in the lateral direction.  In these safety margins, the turbulent kinetic energy was set to zero.  This test case confirmed that the Young-Van-Vliet filter does not require safety margins. For case C-CC-double, the seeding area was increased to span two pitches instead of one. This was performed in order to confirm that the acoustic radiation of all blades can be assumed to be uncorrelated.



\subsection{Setup of CFD computation}

TRACE was used to perform the q3D URANS simulation at 50\% of the stator height of the baseline SDT fan. Periodic boundary conditions were used for a reduced duct consisting of two rotor blades and five stator vanes.  The investigated operating condition was approach.  However, the static pressure at the stator outlet needed to be slightly modified in order to avoid a separation of flow.  

In the stator domain, unsteady solutions of the first 15 harmonics of the rotor blade passing frequency were used to accurately reconstruct the rotor wakes in the absolute frame of reference.  The number of harmonics required mainly depends on the flow gradients in the wake.  The higher the gradients, the more harmonics are required for an accurate representation of the wake.  At mean flow, i.e. at the 0th harmonic, the rotor wakes were averaged via flux averaging at the mixing plane between the moving frame of reference in the rotor block and the absolute frame of reference in the stator block.  When constant turbulence and flow characteristics were considered, the 0th harmonic of the URANS computation rather than a RANS computation was used in order to guarantee consistency.

The Hellsten explicit algebraic Reynolds stress model as implemented by \citet{franke_turbulence_2010} was used to sufficiently reproduce the turbulent characteristics in the leading edge and wake regions. The turbulence in the blade boundary layer was fully resolved by the used mesh.

The spatial discretization was done via a MUSCL (Monotonic Upstream Scheme for Conservation Laws) method of second order accuracy based on Fromm's scheme, while the time discretization was done using an Euler Backward scheme of second order accuracy.  The grid contained nearly 900,000 cells. 



\begin{figure}
\centering
\includegraphics[width=\textwidth,trim=4 4 4 4,clip]{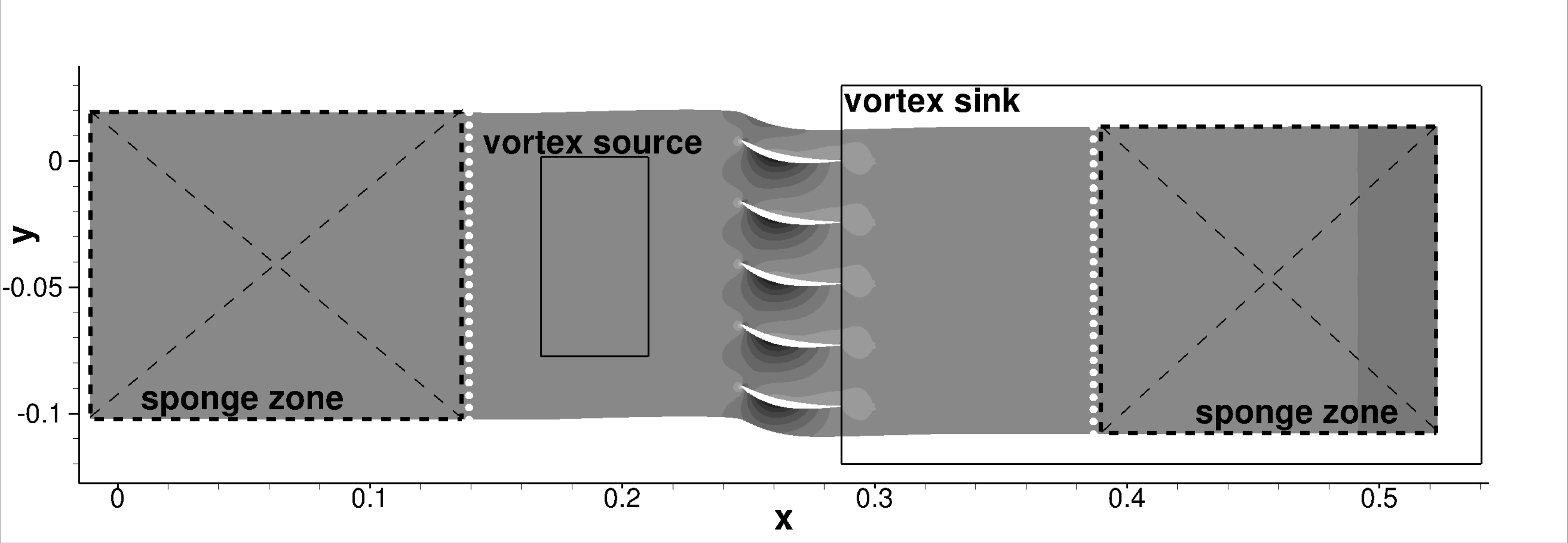}
\caption{CAA setup for five stator vanes showing the sponge zones, the vortex source, and the vortex sink.  The white dots indicate sensor positions and the contour shows the pressure of the background flow, i.e. the mean flow. \label{fig:CAA_Setup} }
\end{figure}

\subsection{fRPM patch generation and setup of CAA computation} \label{subsec:SetupFRPMCAA}
Figure~\ref{fig:CAA_Setup} displays the typical CAA setup for the computation of fan broadband noise with the fRPM method for a reduced duct containing only five stator vanes.  The setup for the full cascade case C-CC-55 was done analogously.  The \textit{vortex source} was generated via a fRPM patch. In the \textit{vortex sink} region, the vorticity is filtered out.  \textit{Sponge zones} were applied at the in- and outlet boundaries of the CAA domain and the white dots indicate the positions of the used \textit{sensors}.  In this subsection, the CAA mesh generation will discussed before examining the italic keywords in detail.

For high-order spatial discretization schemes, a high grid quality of the CAA domain is essential.  The used grid resolution was determined by two factors: acoustics and turbulence.  The CAA mesh was designed to enable the propagation of sound waves up to a frequency $f$ of \SI{20}{kHz} without significant dissipation.  The lowest acoustic grid resolution was chosen to be 10 points per wavelength (PPW), which is approximately twice the theoretical resolution limit of 5.4 PPW for a DRP scheme \cite{de_roeck_overview_2004}. This leads to the following acoustic grid resolution: $dx_f\leq\frac{\lambda}{\text{PPW}}=\SI{1.7e-3}{m}$.  In regions of the CAA domain, where synthesized turbulence is injected into and convected in the domain, a smaller grid resolution is required.  To fully resolve this turbulence, vortical structures have to be resolved with at least 4 points per length scale.  In this case,  the determined grid resolution was given from the smallest Gaussian length scale needed to reconstruct a von \Karman turbulence spectrum through a superposition of ten analytically weighted Gaussian spectra. The analytical weighting function and these best-practice rules are derived by \citet{wohlbrandt_analytical_2016} to realize smooth von \Karman spectra.
The von \Karman turbulence spectrum was determined from experimental data of \citet{podboy_steady_2003} to realize an integral turbulent length scale of $\Lambda\approx\SI{5.1e-3}{m}$ at 50\% of the stator height at the stator LE.  The resulting turbulent grid resolution was  $dx_v\leq\SI{0.51e-3}{m}$.  Furthermore, the boundary layer of the background flow was considered but not fully resolved in the CAA domain. 
Previous studies have shown that artificial sound is generated at a blunt stator trailing edge (TE) if the boundary layer is neglected.  The vortices that move along the blade surface create sound when interacting with a blunt TE.  This is not problematic for pointed TEs but for blunt TEs as is the case for the baseline SDT stator.  The boundary layer prevents the direct interaction between the stator TE and the vortices as it pushes the vortices further away from the blade surface.  The final CAA grid contained 43,324 grid cells per stator passage.     

\begin{figure}
\centering
		\subfigure[Initial patch\label{fig:PatchIni}]{
			\includegraphics[width=0.25\textwidth,trim=0 6 4 4,clip]{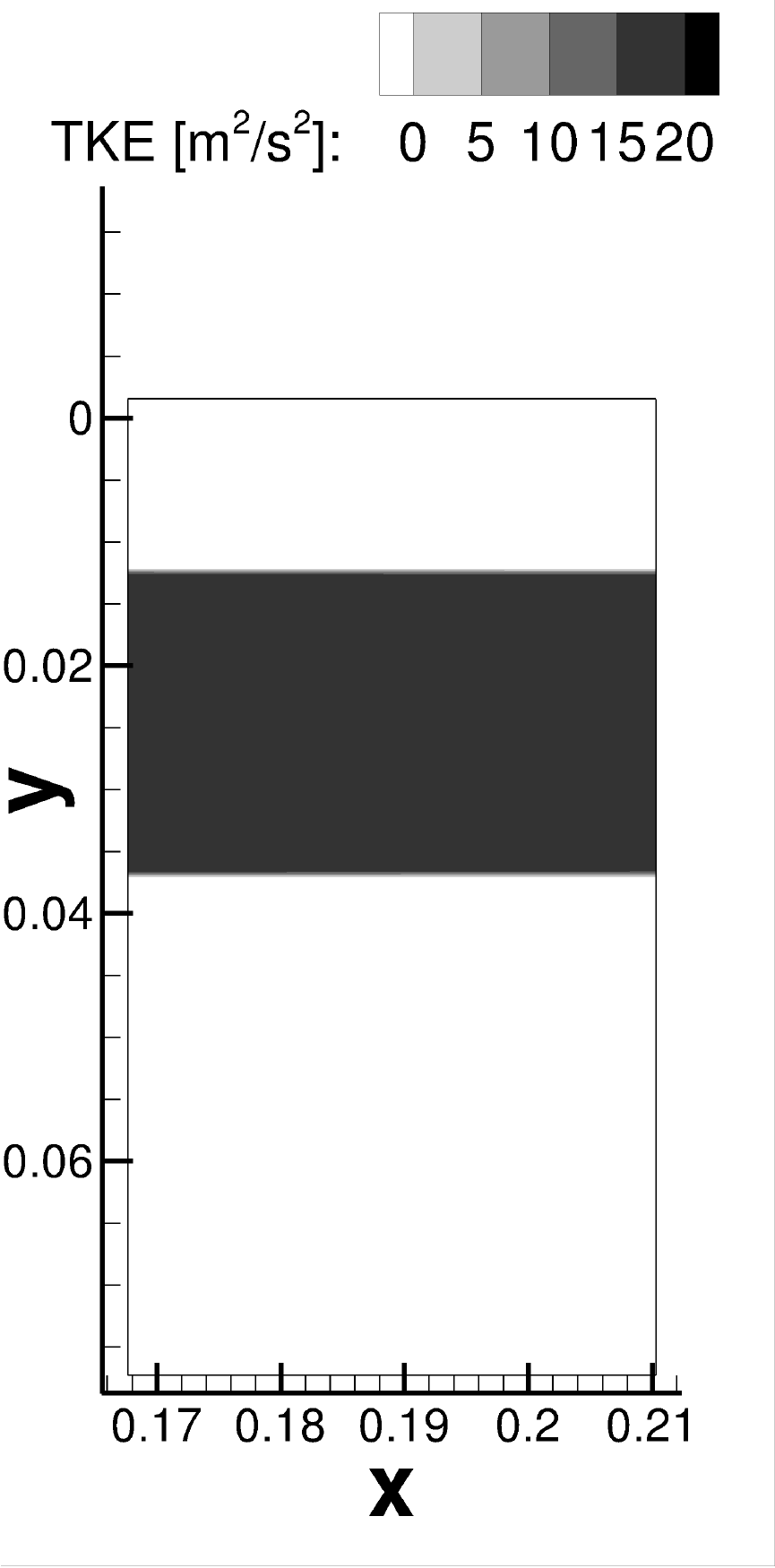}}
		\subfigure[Patch with reduced safety margins\label{fig:PatchRedDamp}]{
			\includegraphics[width=0.25\textwidth,trim=0 6 4 4,clip]{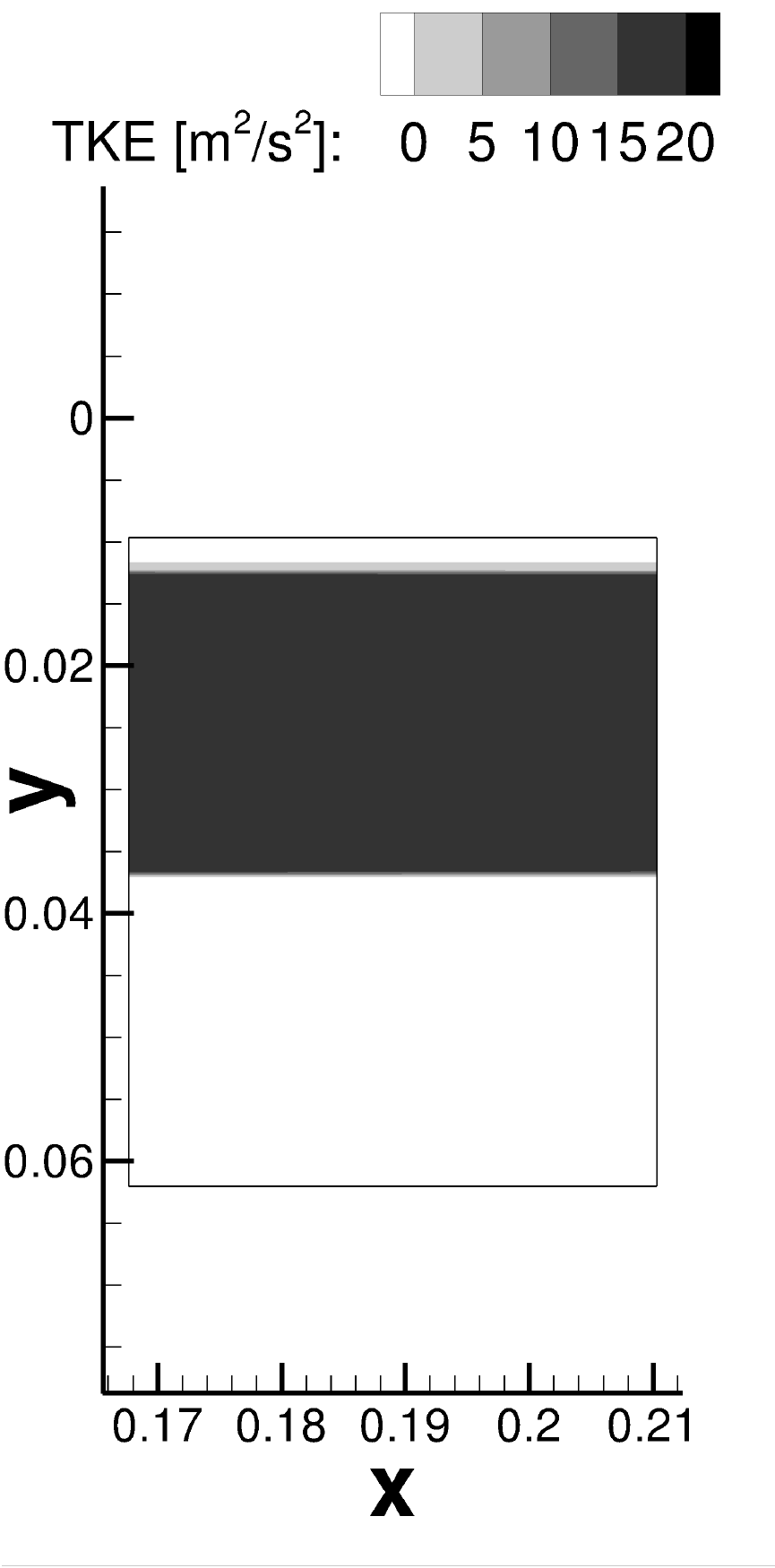}}
		\subfigure[Patch spanning two pitches\label{fig:Patch2Blades}]{
			\includegraphics[width=0.25\textwidth,trim=0 6 4 4,clip]{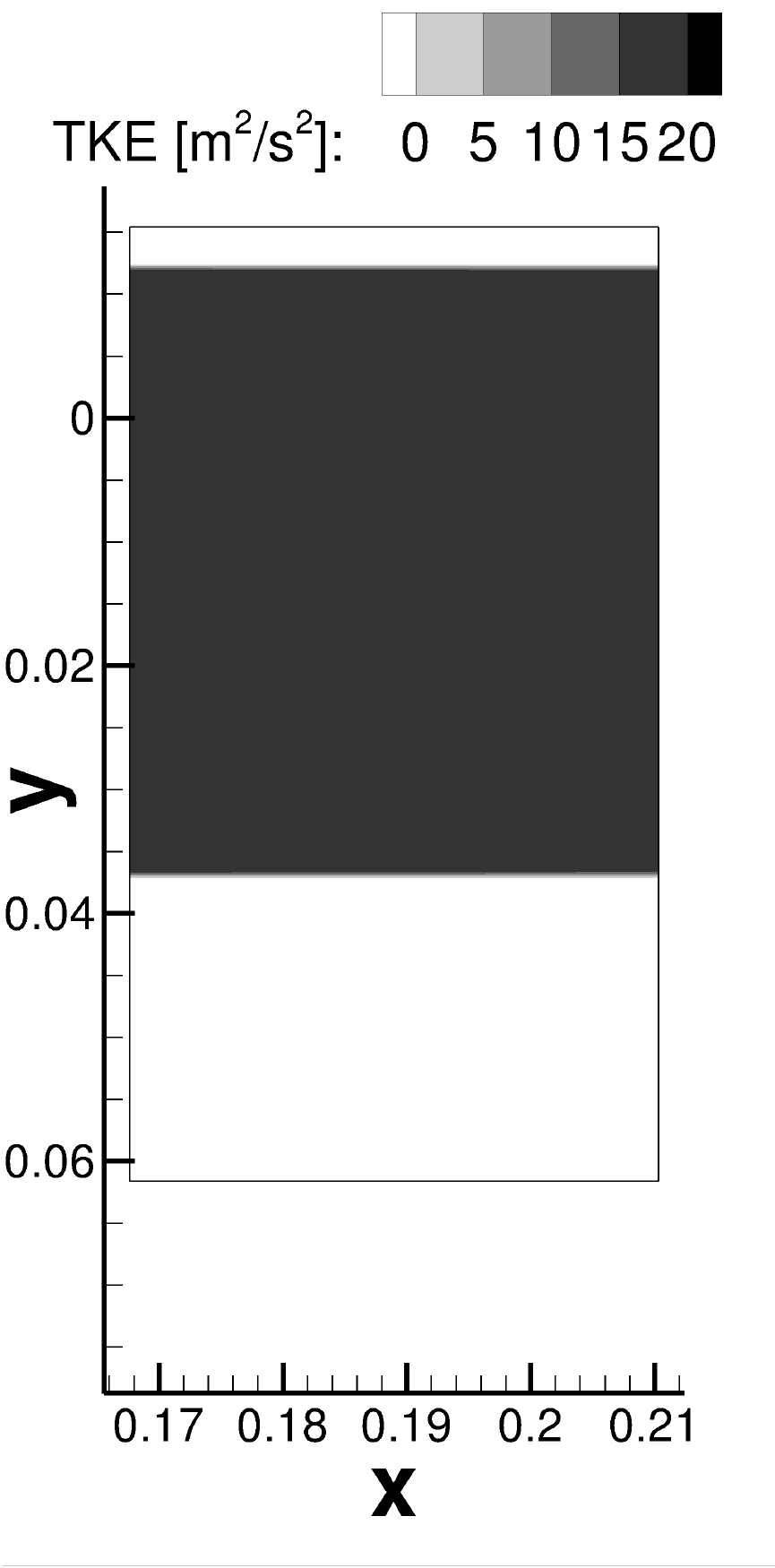}}
	\caption{Used patches for C-CC configuration.  The contour shows the TKE of the mean flow.\label{fig:Patches}}
\end{figure}

\begin{description}
\item{The \textit{vortex source}} produces the synthesized turbulence using inputs from the (U)RANS computation for TKE, TLS, and mean flow and is coupled into the domain using the LEE-relaxation method.  The LE of the source region was located approximately two chord lengths upstream of the stator LE.  The mesh resolution was kept constant to the CAA mesh resolution and five particles per grid cell were used.  As mentioned in the preceding paragraph, a von \Karman spectrum was chosen to generate the synthetic turbulence as a von \Karman spectrum closely emulates realistic turbulence spectra found in turbomachinery. 

Assuming uncorrelated vanes, the RSI is only simulated for a single vane.  All other vanes ensure the correct acoustic radiation only.  Figure~\ref{fig:PatchIni} shows the initial patch used for the primary test matrix.  The contour shows the TKE levels.  The region of the patch containing TKE spans exactly one pitch in order to interact with exactly one stator vane.  The rest of the patch area does not contribute to the synthesized turbulence but acts as safety margin and accounts for lateral convection.  Since the Young-Van-Vliet filter does theoretically not require safety margin as it works on a ghost layer instead, the safety margin was significantly reduced for configuration C-CC-red [see Figure~\ref{fig:PatchRedDamp}].  For configuration C-CC-double, the patch spanned two pitches and has no safety margins [see Figure~\ref{fig:Patch2Blades}].

\item{The \textit{vortex sink}} cancels out vorticity downstream of the stator trailing edge utilizing the LEE-relaxation method by \citet{ewert_linear-_2014}.  The vortices in the CAA domain downstream of the stator TE interact with the stator wakes creating hydrodynamic pressure disturbances.  To get a clean acoustic pressure signal at the sensor position, the vortices were removed from the domain and only acoustic perturbations, which are of interest, remained.  This was done by setting the target vorticity $\bfm\Omega^\text{ref}$ to zero in the whole \textit{vortex sink} region as marked in Fig.~\ref{fig:CAA_Setup} and as alluded to in subsection~\ref{sec:coupling}.  

\item{The \textit{sponge zones}} were needed to avoid reflections at the in- and outlet boundary conditions of the CAA domain.  The sponges were 85 cells deep and a small cell stretching in axial direction - not exceeding a value of 1.1 -  was also introduced.

\item{\textit{Sensors}} were equally spaced up- and downstream of the source region to compute the emitted sound power $P$ of an equivalent 3D duct from the 2D cascade, as derived in the Appendix~\ref{sec:PWL}. As 2D turbulence is generated by the 2D RPM method the correction in Eq.~\ref{eq:2D3DGeschwSpektrum} was used, assuming that the lateral velocity component is the major noise source. The integration surface $S$ resulted from a line of the considered 2D cascades discretized in the y-direction by 275 probes for the full-cascade configuration and 25 probes for the five-vane configuration. As the turbulence only impinged on a single vane, the sound power was multiplied by the number of vanes to get the overall sound power.
\end{description}
The setup for the different configurations remained exactly the same.  Only the number of considered harmonics for the TKE, TLS, and mean flow varied.  Periodic variables were realized by 15 harmonics, while constant variables were realized by the 0th harmonic, i. e. the mean variables.  The computation for the C-CC test case lasted approximately four days on five Intel(R) Xeon(R) CPU E5-2630 v3 CPUs.  The same setup for the P-PP took about eight days to compute.  Realizing cyclostationarity is computationally more expensive as summing up the harmonics at each time step takes more time and more information has to be stored.

\section{Results and discussion}\label{sec:results}

In the following section, the results of the test matrix are discussed.  At first, the primary test matrix is examined to study the effects of cyclostationarity.  Next, an analytical test case is analyzed to substantiate and augment the findings of the first subsection.  The comparison to the measurements can be found in subsection~\ref{sec:meas}. In the last two subsections, elemental assumptions made during the design of the primary test matrix are checked: 1.) The authors assumed that results of a simulation on a reduced duct containing only five stator blades are equivalent to results of a simulation on the full duct containing all 55 stator blades. 2.)  The authors assumed that the RSI noise generation mechanism of each blade is uncorrelated to that of all other blades in the duct and that it is therefore sufficient to only study one blade.

\subsection{Analysis of the primary test cases}
\begin{figure}
\centering
\includegraphics[width=0.9\textwidth,trim=4 4 1 1,clip]{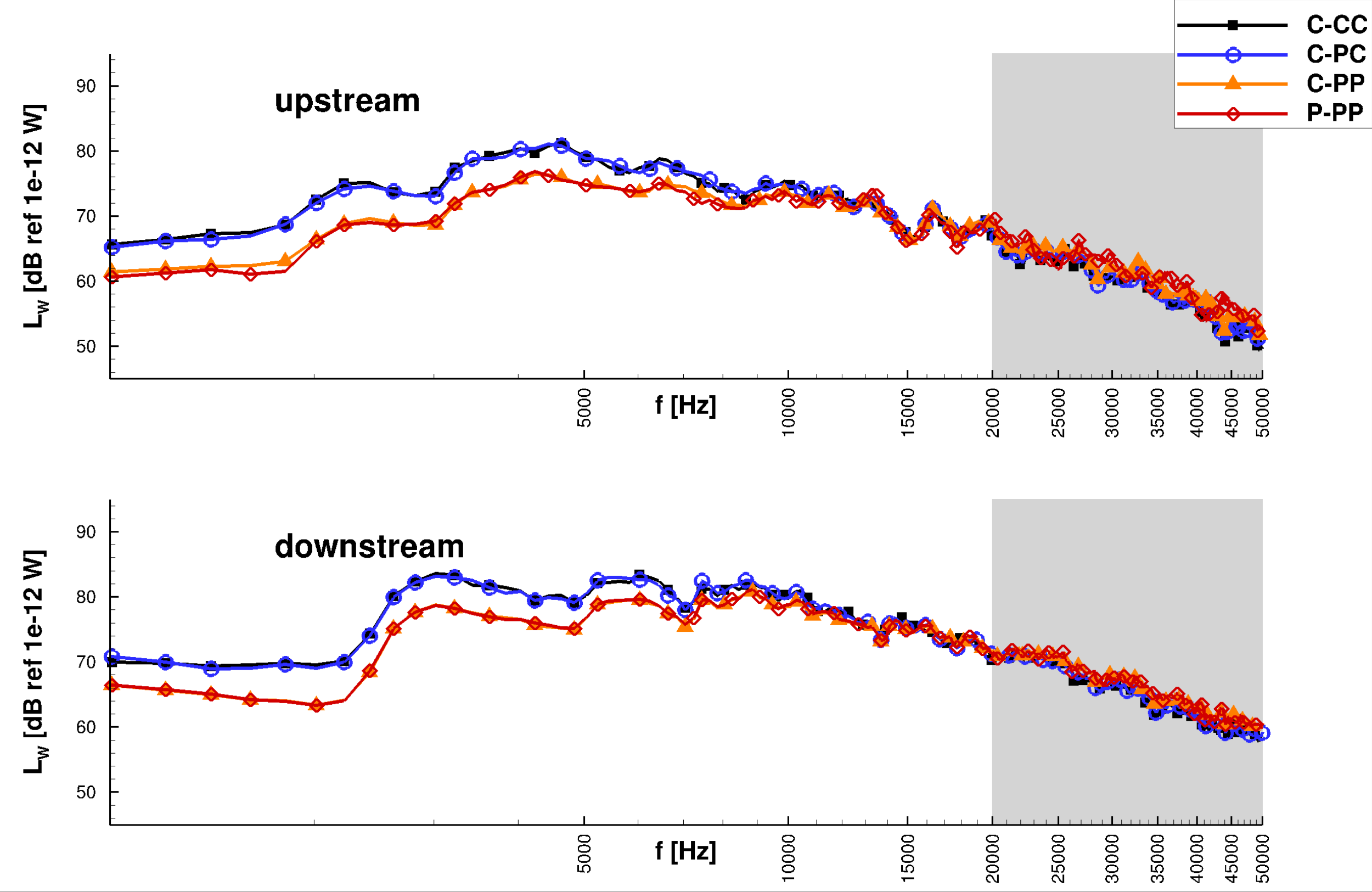} 
\caption{Comparison of sound power spectra of the primary test matrix. The greyed-out regions indicate frequencies outside the range, for which the CAA mesh was designed. \label{fig:PowerSpectra} }
\end{figure}

The primary test matrix was designed to systematically study the effect of cyclostationarity in the turbulence and in the mean flow on the broadband RSI noise.  The study was conducted for a well-known fan, i.e. the NASA SDT fan in its baseline configuration.

In Figure~\ref{fig:PowerSpectra} the sound power level  $L_W$ 
\begin{align}
L_W = 10 \log_{10}\left( P\right/P_\text{ref}) \label{eq:soundPowerLevel}
\end{align}
are shown for all primary test cases with $P$ in Eq.~\ref{eq:P} and $P_\text{ref} = \SI{1e-12}{\watt}$ calculated up- and downstream of the stator.  The grayed-out areas in the graph indicate frequencies outside the range of the target CAA mesh resolution. As the mesh resolution is twice as high as recommended in the literature, results up to \SI{40}{kHz} are deemed to be trustworthy.

When neglecting the periodic nature of the mean flow, the difference seems to be negligible for this configuration as can be concluded when comparing the P-PP and the C-PP primary test cases.  However, when the TLS is taken to be constant over the stator pitch (C-PC) instead of periodic (C-PP), it results in a notable offset in sound power. The offset is large at lower frequencies and disappears at very high frequencies.  Lastly, there is little to no difference in the sound power spectra between the C-CC and C-PC test cases indicating that the cyclostationarity of the TKE does not influence the sound power levels.  To sum up, only the cyclostationarity of the TLS influences the RSI noise for the baseline SDT configuration at approach conditions.

\begin{figure}
\centering
\includegraphics[width=0.9\textwidth,trim=4 4 1 1,clip]{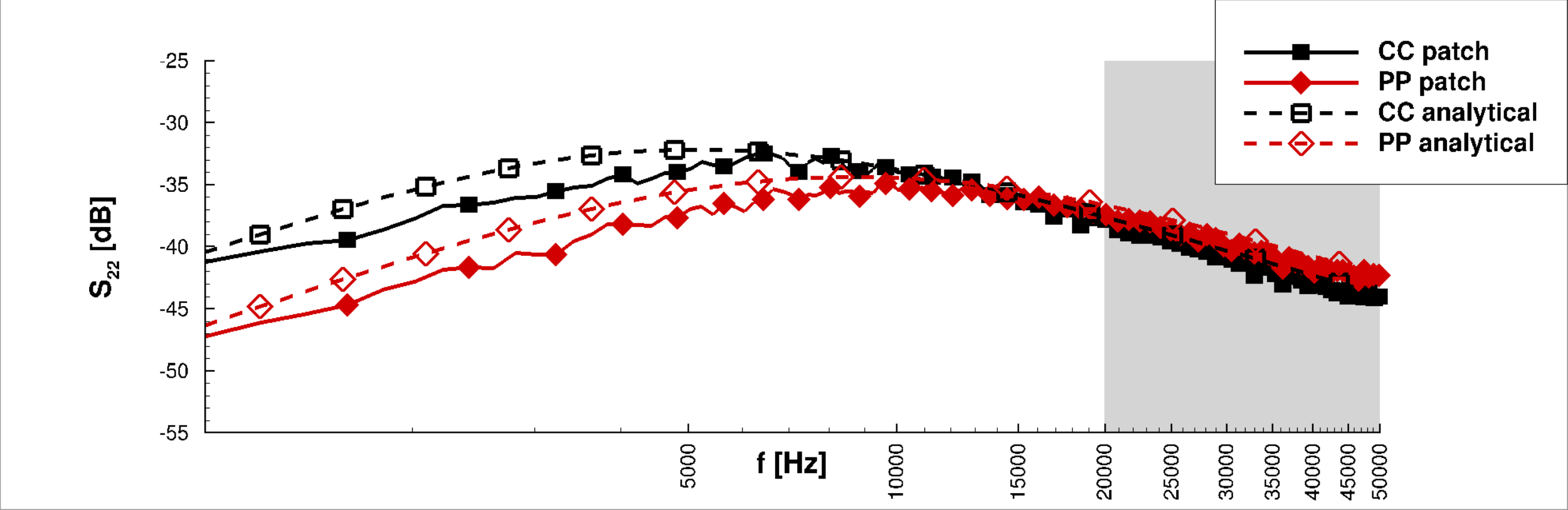} 
\caption{Upwash velocity spectra shown for constant and fully periodic patches.  Comparisons to the respective analytical 2D velocity spectra computed with Eq.~\ref{eq:E22_Karman_2D} and Eq.~\ref{eq:Sii} are shown. \label{fig:PatchSpectra} }
\end{figure}
\subsection{Alternative averaging to realize correct TLS}
In order to examine the effect of the TLS in more detail, the authors took a closer look at the synthesized turbulence in the fRPM vortex source patch.  In Figure~\ref{fig:PatchSpectra}, the lateral 2D velocity frequency spectrum $S^{2D}_{22}(f)$, which is most relevant for the broadband noise generation at the stator LE, realized by the patches for configurations C-CC and C-PP along with their respective analytical solutions are shown. For the constant case, the analytical solutions are calculated using the Eq.~\ref{eq:E22_Karman_2D} and Eq.~\ref{eq:Sii} for the 2D von \Karman turbulence spectrum. For this purpose the turbulent kinetic energy, the turbulent specific dissipation rate and the flow velocity are circumferentially averaged: 
\begin{align}\label{eq:LambdaC}
k_t^C = \frac{1}{2\pi}\int\limits_0^{2\pi} k_t(\vartheta)\abl \vartheta, &&
\omega_t^C = \frac{1}{2\pi}\int\limits_0^{2\pi} \omega_t(\vartheta)\abl \vartheta, &&
u_0^C = \frac{1}{2\pi}\int\limits_0^{2\pi} u_0(\vartheta)\abl \vartheta.
\end{align}
The TLS was determined by Eq.~\ref{eq:Lambda}. For the fully periodic case (P-PP), the analytical velocity spectrum results from integrating over the analytical spectra at each circumferential point of the downstream patch border:
\begin{align}\label{eq:Spii}
S^\text{P}_{ii}(f) = \frac{1}{2\pi}\int\limits_0^{2\pi}S_{ii}(f,\vartheta) \abl \vartheta.
\end{align}  
The TLS is then determined by fitting the averaged velocity frequency spectrum to a standard von \Karman velocity spectrum.  

It can be noted that the numerical spectra match the respective analytical spectra well, particularly at high frequencies.  There is an offset at lower frequencies, which may indicate that the turbulence is lacking energy at those frequencies.  The pronounced offset between the fully periodic and fully constant spectra is most significant.  When a fit is performed for the PP spectra,  the averaged TLS was $\SI{0.0023}{m}$.  In contrast, the averaged TLS for the CC spectra was $\SI{0.0039}{m}$, while the averaged TKE and flow velocity were nearly equivalent.  These findings indicate that the method used for averaging plays a significant role in determining the turbulence characteristics.  Averaging the TKE, TLS, and mean flow over the circumference before calculating velocity frequency spectra (CC) gives different results than calculating spectra for each TKE, TLS, and mean flow before averaging the spectra over the circumference (PP).  
\begin{figure}
\centering
\includegraphics[width=0.9\textwidth,trim=4 4 1 1,clip]{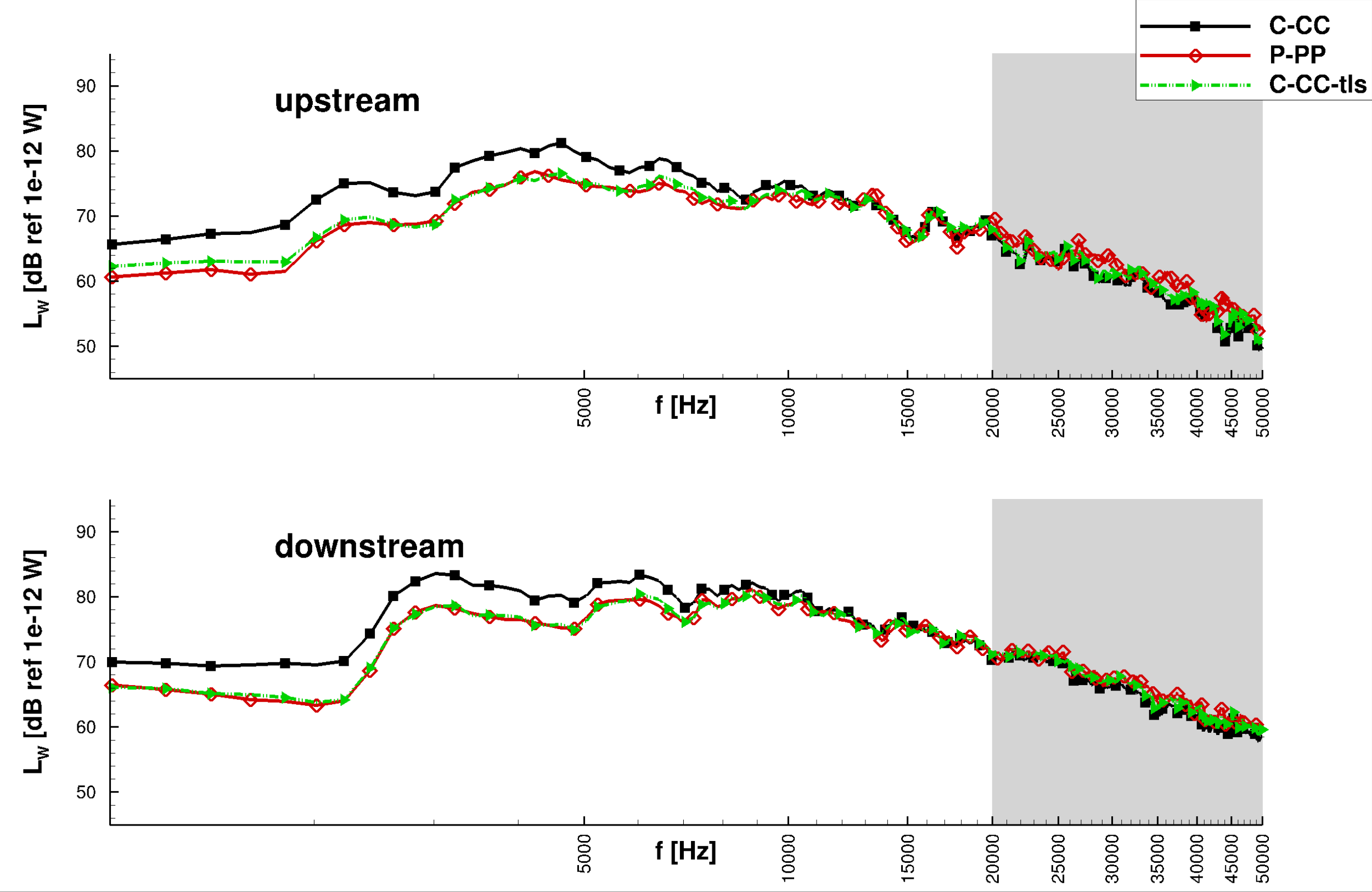} 
\caption{C-CC test case with adjusted turbulent length scale compared to primary test cases C-CC and P-PP \label{fig:PowerSpectra_ls} }
\end{figure}

To further test these findings, the C-CC-tls configuration was simulated.  Instead of using an averaged TLS of $\SI{0.0039}{m}$, a constant TLS of $\SI{0.0023}{m}$ as determined by the PP upwash velocity frequency spectrum was imposed onto the patch.  If the hypothesis that the averaging technique is essential when considering cyclostationary processes is true, the simulation with the modified TLS should produce the same results as the fully periodic simulation (P-PP).  In fact, this is confirmed by the power spectra shown in Figure~\ref{fig:PowerSpectra_ls}.                 

\subsection{Impact of cyclostationarity for an analytic test case}
\begin{figure}[htb]
\parbox{0.39\textwidth}{
\centering
\includegraphics[width=0.39\textwidth,trim=4 4 1 1,clip]{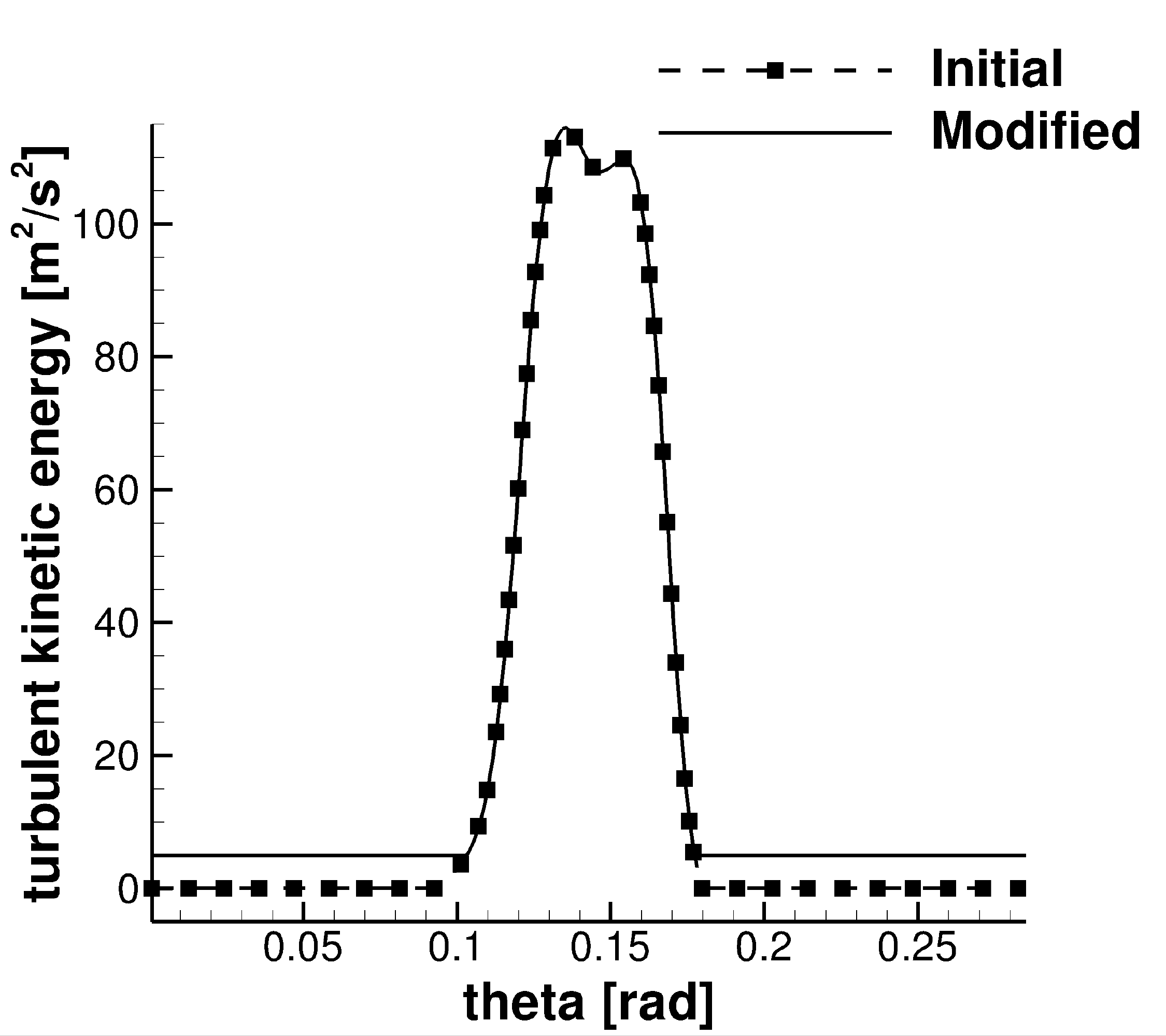} 
\caption{Turbulent kinetic energy of extracted wake.  Background turbulence modified to demonstrate impact on cyclostationarity.\label{fig:TKEWake_analytical} }
}\hfill
\parbox{0.59\textwidth}{
\includegraphics[width=0.59\textwidth,trim=4 4 1 1,clip]{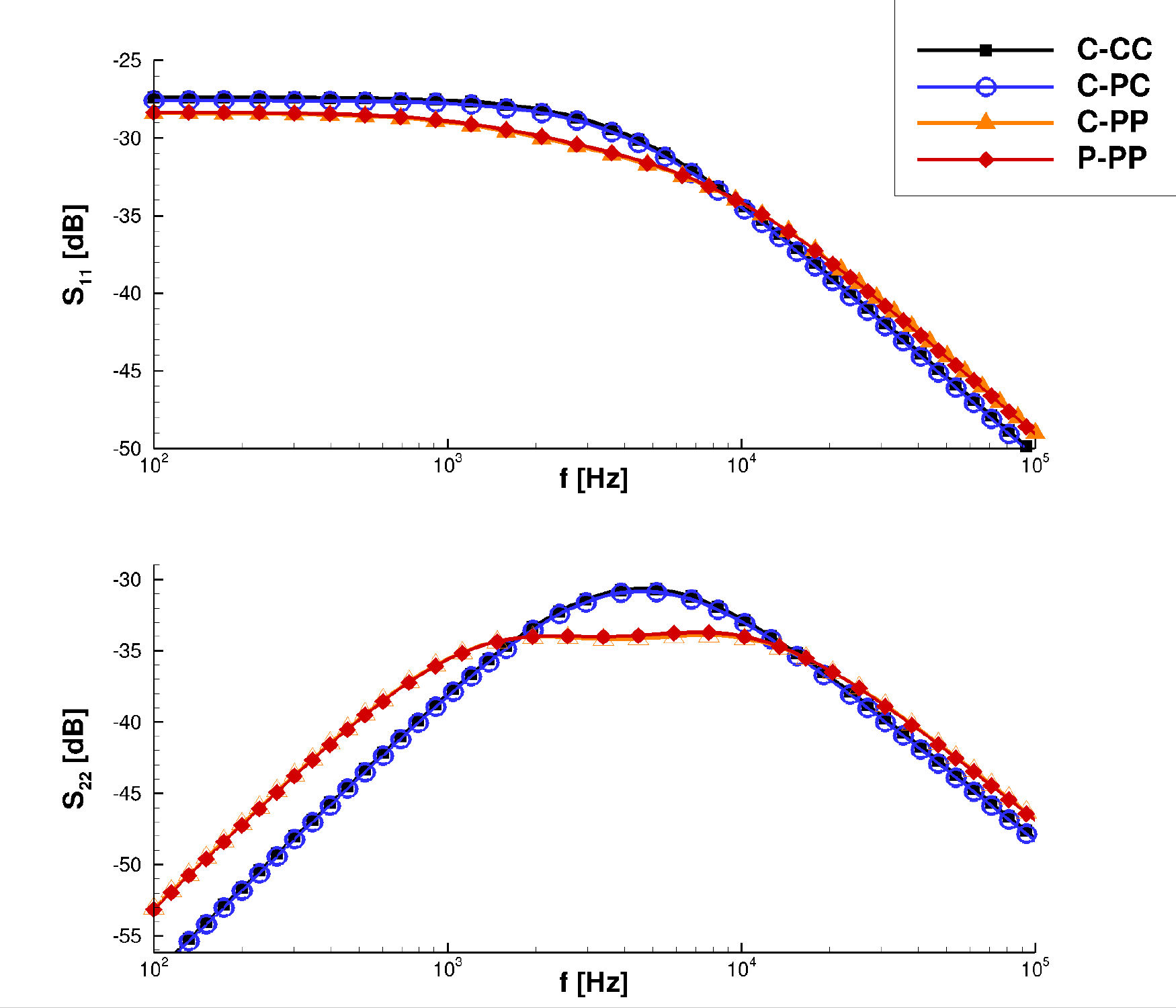} 
\caption{Analytical 2D velocity spectra calculated for modified wake using different averaging techniques. \label{fig:TKESpectra_analytical} }
}
\end{figure}
In the previous subsection, results of the fully periodic test case were reproduced using a fully constant simulation with a TLS determined by a different circumferential averaging technique.  This finding is convenient, since it allows for the use of a less computationally expensive technique granted that the TLS is chosen accordingly. However, for all we know, this finding is only valid for this particular fan configuration at this particular operating point.  To test this, the authors aimed at finding an analytical test case, for which the P-PP case cannot be correctly reproduced by a C-CC case.


In order to achieve this, we slightly manipulated the initial TKE.  Figure~\ref{fig:TKEWake_analytical} shows the TKE of two wakes over one rotor passage.  The initial wake was extracted from the (U)RANS simulation.  The TKE of the background turbulence was set to a constant value to clearly show the difference between the initial and modified cases.  The constant TKE value of the background turbulence is equivalent to the extracted, slightly fluctuating values and results in the same velocity frequency spectra as before.  For the modified case, the TKE in the wake remained the same.  Only the turbulence intensity of the background turbulence was increased from 0.1\% to 1\%.  All other variables remained unchanged.  The resulting velocity frequency spectra by applying the different averaging techniques for the modified wake are shown in Figure~\ref{fig:TKESpectra_analytical}.  As observed in the previous subsection, the periodic variation of the TKE and the mean flow have no impact.  Though the difference in the shape of the velocity frequency spectra due to the periodicity of the TLS is compelling, particularly when considering the direction perpendicular to the flow.  The fully periodic velocity frequency spectrum now exhibits two pronounced bumps instead of just one.  The bump at the lower frequency is due the background turbulence, while the bump at a higher frequency can be attributed to the wake turbulence.  For this hypothetical case, test cases using a constant value for the TLS will never be able to reproduce the shape of the spectrum correctly.

The analytical case with an increased background turbulence intensity proved that there are cases for which the best averaging technique is of no use and cyclostationarity must be simulated in order to synthesize realistic turbulence.

\subsection{Comparison to measurements}
\label{sec:meas}
\begin{figure}
\centering
\includegraphics[width=0.9\textwidth,trim=4 4 1 1,clip]{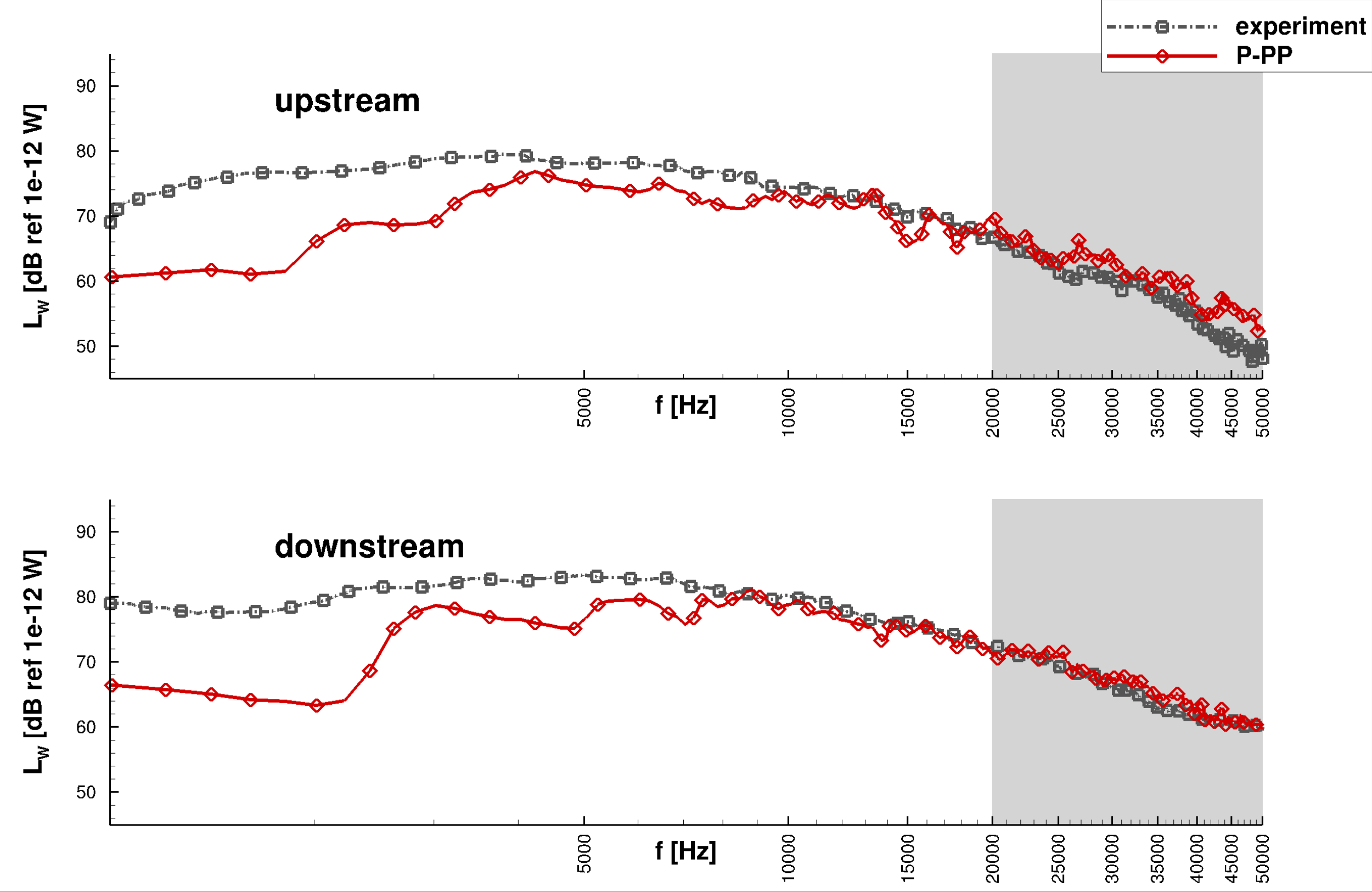} 
\caption{Comparison of numerical and experimental sound power level spectra. \label{fig:PowerSpectraExp} }
\end{figure}
The fully periodic case (P-PP) most accurately reproduces the actual physics.  
For this case the sound power levels are compared to the measurements\footnote{The rotor-stator noise contribution was obtained by subtracting the rotor-alone results from the overall noise results (rotor+stator).} in Figure~\ref{fig:PowerSpectraExp}. The overall trend and levels are reproduced.  While the experimental data differs from the numerical results up- and downstream of the fan in regions below \SI{10}{kHz}, the high-frequency fall off is well predicted. An under-prediction of the sound power at lower frequencies has also been shown by \citet{nallasamy_computation_2005}, who used a RANS-informed, analytical method, at both inlet (upstream) and exhaust (downstream). 

The differences can be explained twofold: (1) The measurements may have contributions from additional noise sources (e.g. rotor trailing edge, jet noise).
(2) Some simplifications and assumptions were made in the numerical hybrid approach. The TKE, TLS, and mean flow were taken from a (U)RANS simulation and a locally isotropic von \Karman spectrum was assumed.  The data was used "as is" and no adjustments were made to achieve a better agreement with experimental results.  Additionally, the used approach is two-dimensional and can only simulate broadband noise resulting from the interaction of turbulence with the blade surfaces.  Any other sound sources that may have been captured by the measurements cannot be considered by this approach.  The method also neglects the rotor, i.e. transmission losses or reflections at the rotor are disregarded.  

\subsection{Comparison of full and reduced duct computations}
\begin{figure}
\centering
\includegraphics[width=0.9\textwidth,trim=4 4 1 1,clip]{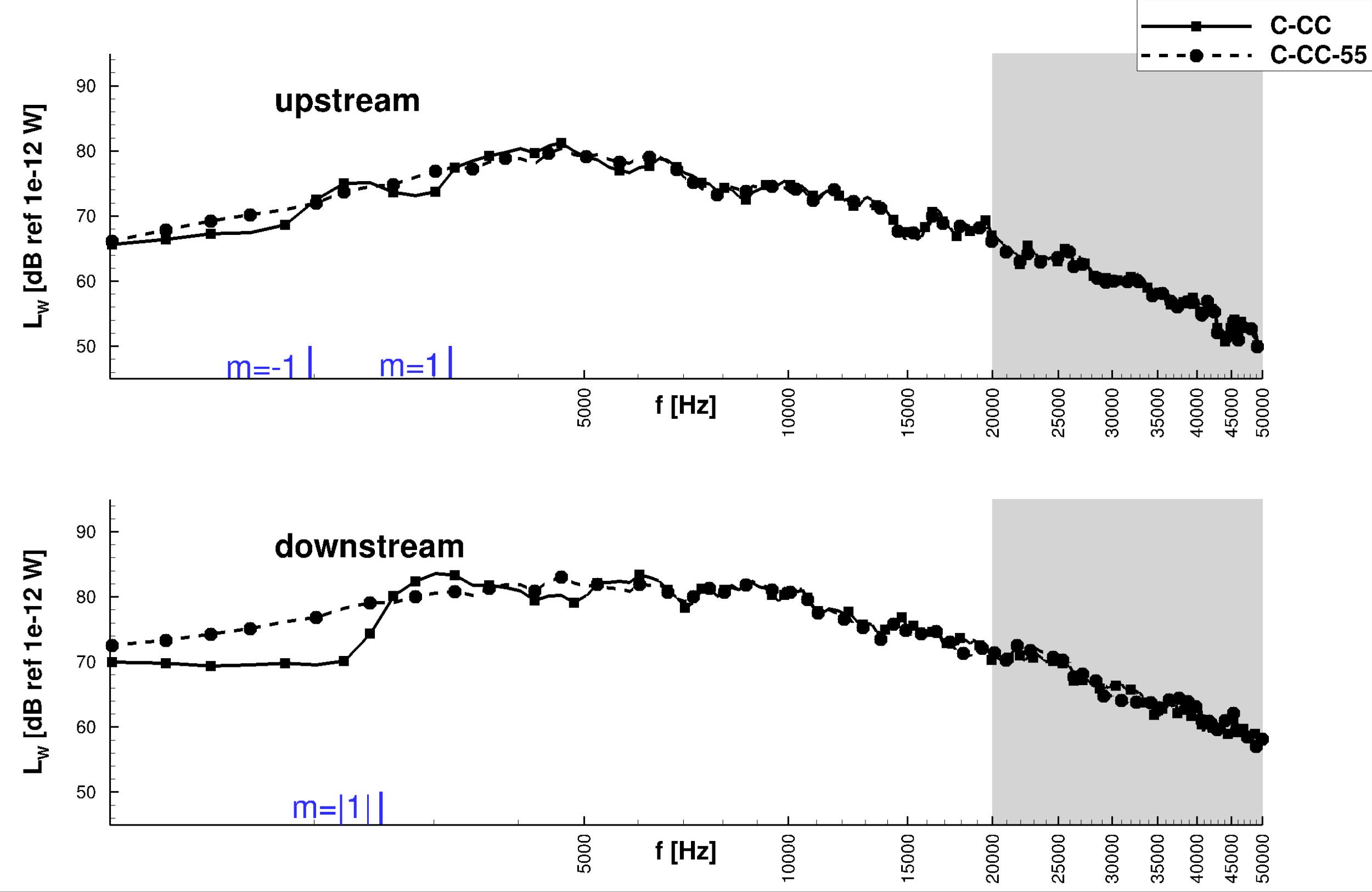} 
\caption{Comparison of the reduced and full cascade using the C-CC configuration.  Cut-on frequencies for azimuthal modes $m=1$ and $m=-1$ are shown. \label{fig:PowerSpectraFullRed} }
\end{figure}
The first elemental assumption that the authors made in the design of the primary test matrix was to assume that simulating a reduced, periodic duct is equivalent to simulating the full duct.  To test this assumption, the C-CC-55 simulation was performed containing all 55 stator vanes.  The resulting narrow-band sound power spectra are shown in Figure~\ref{fig:PowerSpectraFullRed}.  The power spectra are, in fact, nearly equivalent.  Only at lower frequencies, the power spectra of the full cascade is smooth while there are peaks in the power spectra of the reduced cascade. Peaks in fan power spectra are often indicative of where a new acoustic mode of azimuthal order $m$ suddenly becomes cut-on.  The equations for calculating cut-on frequencies of acoustic modes are listed in the Appendix~\ref{app:cuton}.  Aside from the azimuthal mode order, the cut-on frequency depends on the flow speeds and the geometry.  In this case, the duct geometries differ:  The reduced duct with only five stator blades has a smaller circumference than the full duct. For these cases, the relevant Mach numbers are: 
\begin{align*}
	\text{upstream} &&&M_x = 0.40,	&& M_y =-0.21,\text{ and}\\
	\text{downstream} &&&M_x = 0.44,	&& M_y = 0.00.
\end{align*} 
The resulting characteristic frequencies for the first azimuthal mode orders are:

\begin{minipage}[t]{0.5\textwidth}
\begin{itemize}
	\item 5-vane configuration 
	\begin{itemize}
		\item upstream
		\begin{itemize}
			\item[] $f_{m=-1} = 3169.6$~Hz, 
			\item[] $f_{m=+1} = 1968.7$~Hz, and
		\end{itemize}
		\item downstream 
		\begin{itemize}
			\item[] $f_{|m|=1} = 2508.2$~Hz,
		\end{itemize}
	\end{itemize}
\end{itemize}
\end{minipage}
\begin{minipage}[t]{0.5\textwidth}
\begin{itemize}
	\item 55-vane configuration 
	\begin{itemize}
		\item upstream
		\begin{itemize}
			\item[] $f_{m=-1} = 303.9$~Hz, 
			\item[] $f_{m=+1} = 178.4$~Hz, and
		\end{itemize}
		\item downstream 
		\begin{itemize}
			\item[] $f_{|m|=1} = 237.5$~Hz.
		\end{itemize}
	\end{itemize}
\end{itemize}
\end{minipage}
\vspace{0.5em}

For the full cascade, the cut-on frequencies are very low meaning that the acoustic modes of the first and subsequent azimuthal mode orders are cut-on for most of the frequency range.  The determined cut-on frequencies for the reduced cascade are higher and align well with the peaks in the power spectra - both up- and downstream of the stator [see Figure~\ref{fig:PowerSpectraFullRed}].  Aside from the peaks due to the cut-on frequencies, the assumption of the authors was correct and the reduced duct does reproduce the sound power correctly.

\subsection{Acoustic correlation of vane blades}


The second elemental assumption that the authors made was to assume that the investigated sound generation mechanism of each blade is uncorrelated to that of the other blades.  This therefore allows for investigating the impingement of turbulence on only one blade and for multiplying the determined sound power of one blade by the number of blades to receive the total sound power.  In order to confirm this assumption, configuration C-CC-double was investigated with a patch spanning two pitches and therefore turbulence impinging onto two stator blades.

\begin{figure}
\centering
\includegraphics[width=0.9\textwidth,trim=4 4 1 1,clip]{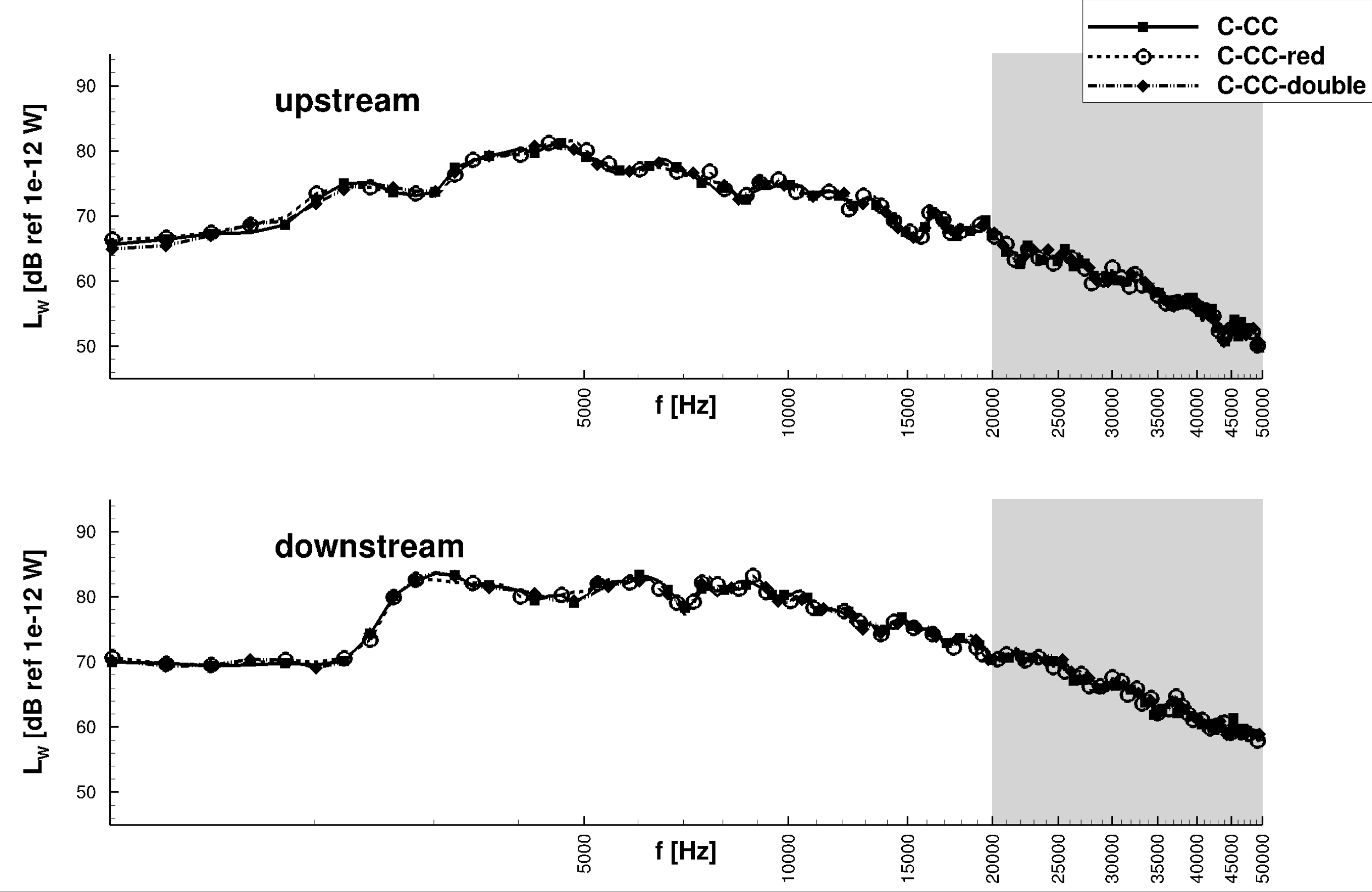}
\caption{Comparison of different patches to check assumption of uncorrelated blades. \label{fig:PowerSpectra_2Pitches} }
\end{figure}
	
An interim step was necessary due to the fact that a patch spanning over two vanes with a lateral safety margin of the same size as the original patch did not fit into the CAA domain for the reduced cascade.  For this interim step, configuration C-CC-red used a patch without a damping zone in lateral direction [see subsection~\ref{subsec:SetupFRPMCAA}].  Figure~\ref{fig:PowerSpectra_2Pitches} shows that the resulting power spectra for the initial patch, used for the primary test matrix, and this patch are identical.  This confirms that the Young-Van-Vliet Filter does, in fact, not need a safety margin in the lateral direction.  Since a different random variable field also had to be used, it also confirms that the solution is independent of the used random variable field.  Since the results were the same, the initial patch spanning one pitch can directly be compared to the patch spanning two pitches.  Both patches yield exactly the same sound power levels [see Figure~\ref{fig:PowerSpectra_2Pitches}].  This confirms that there is no significant acoustic correlation of the vane blades. This result has been anticipated as the TLS is much smaller than the pitch.

\section{Conlusion and Outlook} \label{sec:conclusion}
The cyclostationary stochastic hybrid (CSH) method has been developed based on the RPM method. It enables to account for the effect of turbulence cyclostationarity in fan broadband noise. Since each parameter relevant for the wake turbulence description can be activated individually, it is possible to assess their respective impacts  on broadband RSI noise.  Those parameters are the turbulence kinetic energy, the integral turbulence length scale, and the mean flow.     

The periodic mean flow and periodic turbulence statistics are extracted from a URANS solution that makes use of a Hellsten $k$-$\omega$ turbulence model. The values are passed as Fourier-coefficients to the CAA and fRPM domains. The data can therefore be reconstructed at any arbitrary time step. 

The benefit of using the fRPM-method is a local realization of von \Karman spectra of arbitrary integral length scale and variance in the time domain via analytical weighting of Gaussian spectra. This yields a stable formulation for time variations, even at the steep spatial gradients of TKE and TLS between wake and background turbulence.


One main aspect that motivated this study and the development of CSH method was to investigate which error is made by circumferentially averaging the turbulence and the mean flow before calculating the aeroacoustic blade response.  This averaging technique is commonly used in analytical and most hybrid approaches.  In the investigations presented in this paper, we have observed the following:   
\begin{itemize}
 \item The turbulent length scale is a key parameter. The result depends on the way it is calculated. A simple averaging is not adequate. We propose an alternative method: averaging of the energy/velocity spectrum and fitting that averaged spectrum for the determination of the length scale. With this averaging technique, we can reproduce the results of the fully periodic test case with a constant assumption.
 \item The influence of the cyclostationarity of the TKE and of the mean flow on the sound power levels is negligible for the investigated fan simulation at approach condition.
 \item An analytical investigation shows that the consideration of cyclostationarity is imperative when increasing the background turbulence intensity from 0.1\% to a still realistic 1.0\%. The resulting upwash velocity frequency spectrum differs significantly from a stationary isotropic model spectrum.  This will have a direct impact on the emitted sound.
 \item Furthermore, we have verified that using a reduced number of vanes (5 instead of 55) is a reasonable approach to reduce the computational effort without compromising the results.  Slight differences in the results can only be observed in the low frequency range below the cutoff frequency of the first higher oder mode. Due to the small length scales, it is sufficient to trigger only one blade and consider the signal emitted by all blades to be uncorrelated.
\end{itemize}

The CSH method allows for future, in-depth study of cyclostationary turbulence and will therefore further the understanding of broadband noise generation in fans.  This is especially important for the development and improvement of analytical tools. 

By using stationary spectral analysis to investigate the cyclostationary signals, interesting features are removed from the signal. To look at the intermittency or ''noise events'' of each wake, a cyclostationary analysis~\cite{gardner_cyclostationarity:_2006} has to be utilized. This would highlight variations in the lift coefficient and in the pressure distribution over the time.

The CSH method is currently applied only in two spatial dimensions and uses locally isotropic von \Karman spectra. A 2D-3D correction of the resulting spectra makes it possible to reproduce the measurements fairly well based on numerical data at 50\% duct height at the stator LE. An extension to 3D is contained inherently in the method and will be in focus of future studies.  In addition, modeling anisotropic turbulence by utilizing higher order turbulence models in the URANS solutions is envisioned.


\section{Acknowledgement}
The authors would like to thank Roland Ewert, Jürgen Dierke, and Nils Reiche (DLR) for their advice, enlightening discussions, and valuable time.
Furthermore we would like to acknowledge \citet{envia_panel_2015} for the organization of the fan broadband noise workshop held at the AIAA Aviation conference since 2014. A special thanks goes to Edmane Envia at NASA Glenn Research Center for setting up this test case and providing the extensive data for the SDT fan.


\appendix
\renewcommand\thefigure{\arabic{figure}}  
\section{Appendix}
\subsection{Stream-surface coordinates}
\label{app:mtheta}
The q3D-grid is transformed into a 2D-grid by means of the stream-surface coordinates $m'$ and $\vartheta$.
With the turbo-machine axis defined in positive $x$-direction and the radius defined as $r = \sqrt{y^2+z^2}$, the transformation into stream surface coordinates is given by:
\begin{align}
m' &= \int \sqrt{\abl x^2 +\abl r^2} & \vartheta &= \tan^{-1}(z/y)
\end{align}
In practice this integral is solved by using $m'_0 = 0$ and 
\begin{align}
m'_i &= m'_{i-1} + \frac{2}{r_i+r_{i-1}}
	\sqrt{(r_i+r_{i-1})^2+(x_i+x_{i-1})^2}
\end{align}
This transformation leads to non-dimensional coordinates. 

\subsection{Sound power level}
\label{sec:PWL}
The sound power $P$ is defined as
\begin{align}
P = \int_S n_i I_i \abl S,
\end{align}
with $S$ the integration surface, $n_i$ the surface normal vector, and $I_i$ the net acoustic intensity defined by \citet{Morfey_Sound_1971} as
\begin{align}
I_i = <pu_i> + \frac{u^0_i }{\rho_0 c_0} <pp> + \frac{u^0_i u^0_j}{c_0^2} <pu_j> + \rho_0 u^0_j <u_i u_j >.
\end{align}
Instead of using the rms value indicated by $<ab>$, the cross-spectral density function $S_{ab}$ is used
\begin{align}
<ab> \mapsto  S_{ab}.
\end{align}

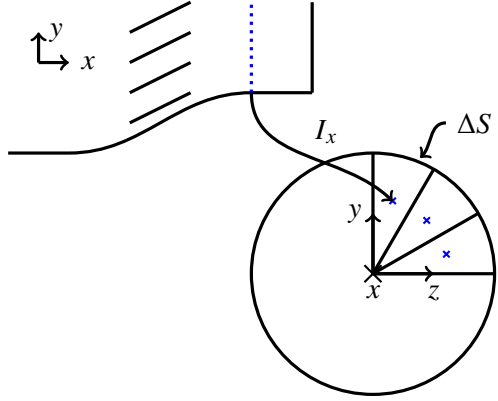
\begin{figure}
	\begin{minipage}{0.7\textwidth}
		\centering
		 
%
%


\tikzset{cross/.style={cross out, draw, 
          minimum size=2*(#1-\pgflinewidth), 
          inner sep=0pt, outer sep=0pt}}
\newcommand{\dist}{3}
\newcommand{\radi}{2}
\newcommand{\coordx}{-5.5}
\newcommand{\coordy}{\dist+0.5}
\newcommand{\coordxD}{0}
\newcommand{\coordyD}{0}

\begin{tikzpicture}[very thick,x=0.8cm,y=0.8cm]
  \draw (-6,\dist-1) -- (-5,\dist-1);
  \draw[out=0, in=180] (-5,\dist-1) to (-2, \dist);
  \draw (-2,\dist) -- (-1,\dist);
  \draw (-1,\dist) -- (-1,\dist+1.5);

  \draw (-4,\dist-0.5) -- (-3,\dist);
  \draw (-4,\dist ) -- (-3,\dist+0.5);
  \draw (-4,\dist+0.5) -- (-3,\dist+1);
  \draw (-4,\dist+1  ) -- (-3,\dist+1.5);

   \draw[dotted,blue] (-2,\dist) -- (-2,\dist+1.5);

   \draw[rotate=0 ] (0,0) -- (\radi,0);
   \draw[rotate=15 ] (0.625*\radi,0) node[cross=2pt,blue,thick] {};
   \draw[rotate=30] (0,0) -- (\radi,0);
   \draw[rotate=45] (0.625*\radi,0) node[cross=2pt,blue,thick] {};
   \draw[rotate=60] (0,0) -- (\radi,0);
   \draw[rotate=75] (0.625*\radi,0) node[cross=2pt,blue,thick] {};
   \draw[rotate=90] (0,0) -- (\radi,0);
   \draw (0,0) circle (\radi);

	\draw[->, out=-90, in = 135] (-2,\dist)  to  node[above right]{$I_x$} (0.2588190451*0.625*\radi,0.96*0.625*\radi);

    \coordinate[<-, label={right:$\Delta S$}] (S) at (0.6*\radi,\radi+0.5);
	\draw[->,out=-180,in=45] (S) to (0.2588190451*3,0.96*\radi);

   \draw [->,rotate around={0:(\coordx,\coordy)}](\coordx,\coordy) -- (\coordx+0.5,\coordy)node[anchor=west] {$x$};
   \draw [->,rotate around={90:(\coordx,\coordy)}](\coordx,\coordy) -- (\coordx+0.5,\coordy)node[anchor=west] {$y$};
   \draw (\coordxD,\coordyD) node[cross=4pt,thick] {};
   \draw (\coordxD,\coordyD) node[anchor=north] {$x$};
   \draw [->,rotate around={90:(\coordxD,\coordyD)}](\coordxD,\coordyD) -- (\coordxD+1,\coordyD)node[anchor=east] {$y$};  
   \draw [->,rotate around={0:(\coordxD,\coordyD)}](\coordxD,\coordyD) -- (\coordxD+1,\coordyD)node[anchor=north] {$z$};  

\end{tikzpicture}
	\end{minipage}\hfill
	\begin{minipage}{0.3\textwidth}
		\caption{Sketch describing the computation of the sound power from a 2D-cascade simulation. In the top left corner, the 2D-cascade and the microphone positions (blue dots) are shown. By rolling up the cascade of radially constant  acoustic intensities on each microphone segment $\Delta S$, a sound power for the whole duct can be determined.\label{fig:powerSketch}}
	\end{minipage}
\end{figure}

For the 2D-cascade the integration surface $\abl S$ is replaced by a discrete segment $\Delta S$. This is shown in Fig.~\ref{fig:powerSketch}. In the upper left corner, the cascade and the microphone positions are shown. Along a line in the $y$-direction, which corresponds to the circumference of the duct $r\theta$, the axial intensity is determined. A correction of 2D to 3D turbulence must be considered assuming the main sound source is the transversal velocity component\footnote{An additional correction for the difference of 2D and 3D propagation does not need to be considered for computations of sound power}. For the von \Karman turbulence spectrum it is given by a quotient $Q_{2D\rightarrow3D}$ of Eq.~\ref{eq:E22_Karman_3D} to Eq.~\ref{eq:E22_Karman_2D}. Assuming that the intensity at position $m$ is representative for a circle segment $\Delta S_m$, the sound power in a comparable duct can be determined by 2D axial intensities of a cascade as
\begin{align}\label{eq:P}
P = Q_{2D\rightarrow3D} \sum\limits_{m=1}^{N_m} I_{x,m} \Delta S_m
\end{align}
with
\begin{align}\label{eq:2D3DGeschwSpektrum}
Q_{2D\rightarrow3D} = \frac{E_{22}^{3D}(k_1)}{E_{22}^{2D}(k1)} = \frac{1}{10}\left(3\hat k_1^{-2}+8\right).
\end{align}
For further definitions refer to the next section.

%

\subsection{Velocity Spectra}\label{app:VelSpectra}
In this investigation, the RPM method generates isotropic turbulence of von Karman shape. For validation, the synthesized spectra are compared to the analytical solution. The velocity one-dimensional wavenumber autospectra in flow direction $\bfm e_1$ and in the direction perpendicular to the flow $\bfm e_2$ are given as:
\begin{align}
E_{11}(k_1) = \frac{u_t^2\Lambda}{\pi} \frac{1}{(1+\hat k_1^2)^{5/6}}  \label{eq:E11_Karman}\\
E_{22}(k_1) = \frac{u_t^2\Lambda}{2 \pi} \frac{1+\frac{8}{3} \hat k_1^2}{(1+\hat k_1^2)^{11/6}} \label{eq:E22_Karman_3D}
\end{align}
with the convective wavenumber $k_1 = \omega/u_0$, the integral length scale (TLS) $\Lambda$, the turbulence velocity variance $u_t$ and the reduced wavenumber $\hat k = k /k_e$ with $k_e = \frac{\sqrt{\pi}\Gamma(5/6)}{\Lambda\Gamma(1/3)}$. For 2D turbulence the lateral velocity one-dimensional wavenumber autospectrum differs and is given as:
\begin{align}\label{eq:E22_Karman_2D}
E_{22}^{2D}(k_1) &= \frac{5 u_t^2\Lambda}{3\pi}  \frac{\hat k_1^2}{\left(1+\hat k_1^2\right)^{11/6}}.
\end{align}
Using the Taylor hypothesis all these can be transformed into frequency space by
\begin{equation}\label{eq:Sii}
S_{ii}(f)=2 E_{ii}(k_1)\frac{2\pi}{u_0}.
\end{equation}

\subsection{Cut-on frequencies}\label{app:cuton}
Often fan spectra calculated with the (linear) wave equations exhibit peaks at frequencies where a new acoustic mode of azimuthal order $m$ and $n$  suddenly becomes cut-on. This is particularly true in the low frequency range where the number of cut-on modes is small. The cut-on frequency in a two dimensional annular duct (infinitely thin) depends on the azimuthal mode order only, the geometry and the flow speed. For flows including swirl, there is a difference for positive and negative azimuthal mode orders $m$. The cut-on frequencies are given by
\begin{align}
f_c(m<0) &= \
\frac{|m| c_0}{N_V s_V} \left(M_y - \sqrt{1-M_x^2}\right), \\
f_c(m>0) &= 
\frac{|m| c_0}{N_V s_V} \left(M_y + \sqrt{1-M_x^2}\right)
\end{align}
with $c_0$ the speed of sound and $M_x$ and $M_y$  the axial and circumferential flow Mach number components, respectively.

\bibliographystyle{model1a-num-names}
\bibliography{bibliography}
\addcontentsline{toc}{section}{References}

\end{document}